\documentclass[journal]{IEEEtran}
\usepackage{algorithm}
\usepackage{algpseudocode}
\usepackage{amsfonts}
\usepackage{bm}
\usepackage{amsmath}
\usepackage{cite}
\usepackage{float}
\usepackage{amssymb}
\usepackage[timestamp,first]{draftcopy}
\usepackage{enumerate}
\usepackage{color,xcolor}
\usepackage{colortbl}
\usepackage{graphicx}
\usepackage{multirow,tabularx}
\usepackage{subfigure}

\makeatletter

\newcommand{\Rmnum}[1]{\expandafter\@slowromancap\romannumeral #1@}
\makeatother

\ifCLASSINFOpdf

\else

\fi

\begin{document}
\title{Practical MIMO-NOMA: Low Complexity \& Capacity-Approaching Solution}

\author{\IEEEauthorblockN{ Yuhao Chi, \emph{Student Member, IEEE,} Lei Liu, \emph{Member, IEEE}, Guanghui Song, \emph{Member, IEEE}, \\ Chau Yuen, \emph{Senior Member, IEEE,} Yong Liang Guan, \emph{Senior Member, IEEE}, and Ying Li, \emph{Member, IEEE}}
\thanks{Y.~Chi and Y.~Li are with the State Key Lab of Integrated Services Networks, Xidian University, Xi'an, 710071, China (e-mail: yhchixidian@gmail.com, yli@mail.xidian.edu.cn).}
\thanks{L. Liu, G. Song, and C. Yuen are with the Singapore University of Technology and Design, Singapore 487372 (e-mail: leiliuxidian@gmail.com, gsong2017@gmail.com, yuenchau@sutd.edu.sg).}
\thanks{Y. L. Guan is with the School of Electrical and Electronic Engineering, Nanyang Technological University, Singapore 639798 (e-mail: eylguan@ntu.edu.sg).}
}
\maketitle

\begin{abstract}
MIMO-NOMA combines {Multiple-Input Multiple-Output} (MIMO) and {Non-Orthogonal Multiple Access} (NOMA), which can address heterogeneous challenges, such as massive connectivity, low latency, and high reliability. In this paper, a practical coded MIMO-NOMA system with capacity-approaching performance as well as low implementation complexity is proposed. Specifically, the employed receiver consists of a multi-user Linear Minimum Mean-Square Error (LMMSE) detector and a bank of single-user message-passing decoders, which decompose the overall signal recovery into distributed low-complexity calculations. An asymptotic extrinsic information transfer analysis is proposed to estimate the performance of iterative receiver, where practical channel codes that match with the LMMSE detector in the iterative decoding perspective are constructed. As a result, the proposed coded MIMO-NOMA system achieves asymptotic performances within $0.2$~dB from the theoretical capacity. Simulation results validate the reliability and robustness of the proposed system in practical settings, including various system loads, iteration numbers, code lengths, and channel conditions.
\end{abstract}


\section{Introduction}

With the popularization of Internet and intelligent technology, the number of communication devices is predicted to reach 40.9 billion in~2020~\cite{IOT0}, which includes new communication scenes, such as machine-to-machine communications \cite{M2M,M2MBook}, Internet of things \cite{IoT}, and vehicle-to-vehicle (V2V) communications \cite{Guo}. Due to the fact that available spectrum resources are limited, orthogonal multiple access technology in the fourth generation (4G) communication system cannot satisfy the massive access demands. As a result, {Non-Orthogonal Multiple Access} (NOMA)~\cite{NOMA,NOMAIot,NOMAma,NOMAsurvey,MONOMA,NOMA_Samith,DingTSP,DingHybrid,Song-ISIT2015,Song-TVT2017,Congbin-2017,Song-MaxSum,Chi,DingGC} emerges to support heavily overloaded communications, which allows multiple users to share the same time and frequency resources. To further improve spectral efficiency and reduce latency, NOMA combining with {Multiple-Input Multiple-Output} (MIMO)~\cite{Tse,DaiMIMO}, termed MIMO-NOMA~\cite{DaiMIMO-NOMAMM,NOMACapacity,Ding2016,DingTWC,DingUser,PrecMIMO-NOMA,SCMA,SCMA2,SongSCMA,SCMAC,
MIMO-SCMA,MIMO-SCMA2,MIMO-SCMA3,Lei2015,Lei2016,Lei20162,MIMOIF,Lei-ICC}, is considered as a key air interface technology in the fifth-generation (5G) communication system \cite{5G,White}.

Theoretical analysis has proved that MIMO-NOMA systems can achieve higher capacity than orthogonal multi-user MIMO systems of 4G \cite{NOMACapacity}. From the perspectives of applications, multiple users in MIMO-NOMA are separated by different transmission powers~\cite{Ding2016,DingTWC,DingUser} or different channel codes~\cite{SCMA,SCMA2,SongSCMA,SCMAC,MIMO-SCMA,MIMO-SCMA2,MIMO-SCMA3,Lei2015,Lei2016,Lei20162,MIMOIF,Lei-ICC}, where the former employs Successive Interference Cancellation (SIC) receiver and the latter relies on a joint iterative multi-user decoding.

In MIMO-NOMA systems with SIC receiver \cite{Ding2016,DingTWC,DingUser}, different users are allocated to different power levels and the SIC receiver decodes and then removes the interference of each user according to a descending order of their channel gains \cite{Tse}. Although the implementation of the power-allocation system is simple, SIC receiver has three inherent problems in practice: (1) error propagation, i.e., residual errors of earlier decoded users still affect the decoding of the later users, (2)
the performance of SIC receiver is sensitive to the accuracy of channel state information (CSI), (3) decoding latencies of the later users might be large especially when the number of users is large.

In MIMO-NOMA systems with joint iterative multi-user decoding~\cite{SCMA,SCMA2,SongSCMA,SCMAC,MIMO-SCMA,MIMO-SCMA2,MIMO-SCMA3}, different users are allocated with different codes before transmission and the joint iterative multi-user decoder detects signals for all users simultaneously. In the works on transmitter design, {Sparse Code Multiple Access} (SCMA), a kind of NOMA, is considered in \cite{SCMA,SongSCMA,SCMA2,SCMAC}, in which multiple users are allocated with different sparse signature codes for user separation.
Works~\cite{SongSCMA} and~\cite{SCMAC} considered codebook design for SCMA based on the criteria of maximum a minimum code distance and mutual information, respectively. However, since the design involves a joint optimization of multiple users' codes, which becomes extremely difficult as the user number is large. Works~\cite{MIMO-SCMA,MIMO-SCMA2,MIMO-SCMA3,Lei2015,Lei2016,Lei20162,MIMOIF} proposed several low-complexity multi-user detection schemes for MIMO-NOMA with near optimal performance, such as the Gaussian message passing detection (GMPD), integer forcing detection, and Linear Minimum Mean-Square Error (LMMSE) detection. Especially, work~\cite{Lei-ICC} proved that the LMMSE detector can achieve the capacity region of MIMO-NOMA system when the employed channel code possesses an EXtrinsic Information Transfer (EXIT) property that perfectly matches with that of the LMMSE detector. Unfortunately, these works did not provide any practical channel code design for MIMO-NOMA with these excellent multi-user detection schemes. This motivates us to design practical channel codes with low-complexity encoding and decoding to achieve this goal.

In this paper, we consider practical code design for uplink MIMO-NOMA system that takes implementation complexity and performance into account at the same time. The major contributions of this paper are summarized as follows.
\begin{enumerate}
  \item An asymptotic analysis is proposed to trace the EXIT property between LMMSE detector and message-passing decoders.
  \item Based on the asymptotic EXIT analysis, we design multi-user encoders such that the message-passing decoders match with the LMMSE detector in iterative decoding perspective. The proposed code has an asymptotic performance with only 0.2 dB from the channel capacity.
  \item We show that the proposed system is robust to various code lengths, iteration numbers, and channel conditions via simulations, and is implementable with a low decoding complexity of $\mathcal{O}((min\{MK^2+K^3, KM^2+M^3\}+K)\tau_{\rm{max}}+K)$, where $K$, $M$, and $\tau_{\rm{max}}$ denote the number of users, receive antennas, and iterative detections respectively.
\end{enumerate}
It should be emphasized that comparing with precoded MIMO-NOMA, where the precoding is generally used for beamforming, power allocation, and user pairing~\cite{DingTWC,DingUser,PrecMIMO-NOMA}, the proposed system does not require instantaneous CSI. Moreover, since our code design aligns with the joint iterative multi-user decoding, a significant coding gain is achieved comparing with the MIMO-NOMA system with a conventional channel code designed for point-to-point channel. Therefore, the proposed system can be an attractive solution for the MIMO-NOMA uplink in 5G communications.

The rest of this paper is organized as follows. In Section II, the model and challenges of MIMO-NOMA are presented. The asymptotic EXIT analysis between LMMSE detector and message-passing decoders is introduced elaborately in Section~III. Section~IV provides a practical coding scheme for MIMO-NOMA system and the analyses of complexity and performance. Section V presents various simulations to validate the reliability and robustness of the proposed MIMO-NOMA system. Finally, Section VI concludes this paper and provides some future works.

\section{System Model and Challenges}

In this section, the system model of uplink coded MIMO-NOMA is presented. Subsequently, the challenges in the designs of transmitter and receiver are discussed, which motivate the overall system design with the goal of achieving capacity-approaching performance at low implementation complexity.
\subsection{System Model}
\begin{figure}
  \centering
  \includegraphics[width=1\columnwidth]{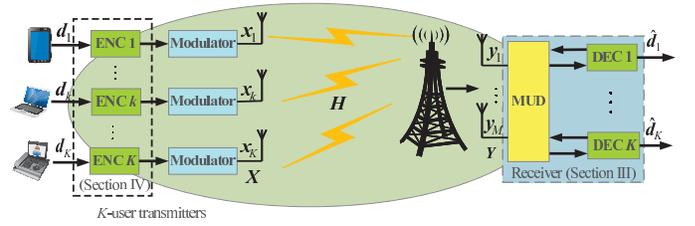}\\
  \caption{Scenario of an uplink coded  MIMO-NOMA system with $K$ single-antenna users and a BS equipped with $M$ antennas.
  ENC, DEC, and MUD denote encoder, decoder, and multi-user detector respectively.}\label{Model}
\end{figure}
Figure~\ref{Model} illustrates an uplink coded MIMO-NOMA system, which includes $K$ single-antenna users and a base station (BS) equipped with $M$ antennas. At the $K$-user transmitters, information sequence $\bm{d}_k$ is encoded by encoder~$k$, $k=1, ..., K$, and comes into the followed modulator. Then, generated symbol sequence $\bm{x}_k$ is transmitted to the channel. Here, we assume that each user has the same transmit rate $R$ and the transmitted power for each user is normalized as $1$.

When all transmitted signals from $K$ users arrive at BS synchronously, received signal $\bm{Y} = [\bm{y}_1,  ..., \bm{y}_M]^{\rm{T}}${\footnote{$[\cdot]^{\rm{T}}$ denotes the transposition of a vector or matrix.}} is
\begin{equation}\label{recv}
\bm{Y} = \bm{H}\bm{X}+\bm{z},
\end{equation}
where $\bm{X}=[\bm{x}_1, ..., \bm{x}_K]^{\rm{T}}$ denotes transmitted signals from $K$ users, $\bm{H}$ is the channel matrix from $K$ users to BS, and $\bm{z}$ is an additive Gaussian noise vector. We assume that $\bm{H}$ is available at the BS but unknown for the $K$-user transmitters.

The goal of the receiver at BS is to recover the signals for all users. As shown in Fig.~\ref{Model}, the employed receiver consists of a multi-user detector (MUD) and a bank of single-user decoders, in which the iterative detection for all users' signals is performed between the MUD and all single-user decoders. Specifically, based on received signal $\bm{Y}$ and \emph{a priori} estimations derived from the decoders, the MUD outputs soft estimations for each transmitted symbol of each user. Based on these estimations from the MUD, a single-user decoding is performed in each decoder, which feeds the output estimations back to the MUD. The whole iterative process will stop when all signals are recovered successfully or the maximum iteration number is reached.

\subsection{Challenges}

To enable the system to achieve capacity-approaching performance at low complexity, we discuss the challenges in the designs of transmitter and receiver, and propose the corresponding solutions.

For the $K$-user transmitters, the challenge is to conceive the encoding scheme for each user so that the $K$-user's messages could be efficiently decoded via the multi-user decoder. Since this multi-user decoding involves a separation of $K$-user's signals from a compound receiver, a sophisticated encoding or preprocessing for each user's signal are required to realize this goal. Although works~\cite{SCMA,SCMA2,SongSCMA,SCMAC,MIMO-SCMA,MIMO-SCMA2,MIMO-SCMA3}~assign user-specific modulation scheme so that the signals of each user could be physically identified at the receiver, alternatively, we apply the same modulation scheme, just the simplest BPSK modulation, for each user, and show that the signals of each user could be well recovered only through our proposed channel coding and decoding. But different from conventional point-to-point codes~\cite{SLin} that are designed specially to overcome channel noises, the proposed code is designed to overcome not only the noise interference but also the multi-user interference from other non-orthogonal users. Therefore, we name the proposed code as multi-user code. The detailed design process for the multi-user code will be discussed below.

For the receiver, the challenge is achieving capacity-approaching performance for signal detection with low complexity. Although each component of the receiver can adopt an optimal algorithm~\cite{MUD}, i.e., Maximum A Posteriori (MAP) algorithm in the MUD and A\;Posteriori Probability (APP) algorithm in the decoders, this optimal solution is severely limited by the prohibitive complexity, which increases exponentially with user number and code length. As a result, we employ an alternative low-complexity LMMSE detection in the MUD and a message-passing decoding in the decoders, which can decompose the overall signal recovery into distributed low-complexity calculations~\cite{LMMSE,MCT,BP-LDPC,FG}. Meanwhile, since the LMMSE detection is proved to be capacity-approaching in the EXIT point of view under iterative decoding~\cite{Lei-ICC}, our objective is to design a practical capacity-approaching code.

\section{Asymptotic Analysis of Iterative Receiver}

In this section, an asymptotic EXIT analysis is proposed to trace the EXIT property between LMMSE detector and message-passing decoders. Based on this asymptotic analysis, the guideline for multi-user code design is provided.

Here, we consider a real-domain system, where each modulator employs BPSK, the elements of channel matrix $\bm{H}$ obey a real Gaussian distribution $\mathcal{N}(0, 1)$, and the elements of channel noise $\bm{z}$ obey a real Gaussian distribution $\mathcal{N}(0, \sigma_n^{2})$. The analysis for complex-domain systems with high-order modulations can be extended accordingly.

\subsection{LMMSE Detection}

The LMMSE detection is used for estimating the transmitted signals of each user. Since the signal estimation for each user is similar, we only focus on the detection of $x_k$ of user $k$.

Based on \emph{a priori} log-likelihood ratio (LLR) $L_{\rm{MUD}}^a(x_k)$ from decoder~$k$, mean $\bar{x}_k$ and variance $v_k$ associated with $x_k$ are
\begin{equation}\nonumber
\bar{x}_k=E\big[x_k|L_{\rm{MUD}}^a(x_k)\big], \quad v_k=E\big[|x_k-\bar{x}_k|^2|L_{\rm{MUD}}^a(x_k)\big],
\end{equation}
where $E[a|b]$ denotes the conditional expectation of  variable $a$ when given variable $b$. Let ${\bar{\bm{X}}}=[\bar{x}_1, ..., \bar{x}_K]^T$ and $\bm{V}_{\bar{\bm{X}}}=diag(v_1, ..., v_K)$\footnote{$diag(v_1, ..., v_K)$ denotes the diagonal matrix with diagonal elements $(v_1, ..., v_K).$}. Based on received signal $\bm{Y}$ in Eq.~(\ref{recv}), \emph{a posterior} estimation ${\hat{\bm{X}}}=[{\hat{x}}_1, ..., {\hat{x}}_{K}]^T$ of LMMSE detector is~\cite{Lei2016}
\begin{align} \label{LMSEOut}
{\hat{\bm{X}}}=\bm{V}_{\hat{\bm{X}}}[\bm{V}_{\bar{\bm{X}}}^{-1}\bar{\bm{X}}+\sigma_n^{-2}{\bm{H}}^T\bm{Y}],
\end{align}
where $\bm{V}_{\hat{\bm{X}}}$$=$$diag(\hat{v}_1, ..., \hat{v}_K)=(\sigma_n^{-2}{\bm{H}}^{{T}}\bm{H}+{V}_{{\bar{\bm{X}}}}^{-1})^{-1}$ denotes the deviation between \emph{a posterior} estimated signal $\hat{\bm{X}}$ and exact signal $\bm{X}$.

According to the message combining rule~\cite{Yuan}, \emph{extrinsic} mean $\hat{x}_k^e$ and variance $v_k^e$ are obtained by excluding $\emph{a priori}$ mean $\bar{x}_k$ and variance $v_k$ from \emph{a posterior} mean $\hat{x}_k$ and variance $\hat{v}_k$:
\begin{equation}\label{EXLMSE}
v_k^e=[\hat{v}_k^{-1}-v_k^{-1}]^{-1}, \quad  \hat{x}_k^e=v_k^e[\frac{\hat{x}_k}{\hat{v}_k}-\frac{\bar{x}_k}{v_k}].
\end{equation}
On the other hand, Eq.~(\ref{LMSEOut}) can be rewritten as
$\hat{\bm{X}}=\bar{\bm{X}}+\bm{V}_{\bar{\bm{X}}}{\bm{H}}^T(\sigma_n^2\bm{I}_M+\bm{H}\bm{V}_{\bar{\bm{X}}}\bm{H}^T)^{-1}(\bm{Y}-\bm{H}\bar{\bm{X}}),$
where $\bm{I}_{M}$ is an $M \times M$ identity matrix. Thus, $\hat{x}_k^e$ can be rewritten as
\begin{align}\label{AWGN}
&\hat{x}_k^e=x_k+\hat{z}_k \\ \nonumber
&=x_k\!+\!\frac{v_k v_k^e}{\hat{v}_k}\bm{h}_k^T(\sigma_n^2\bm{I}_{M}\!+\!\bm{H}V_{\bar{\bm{X}}}\bm{H}^T)^{-1}[\bm{H}\big(\bm{X}_{\setminus k}\!-\!\bar{\bm{X}}_{\setminus k}\big)\!+\!\bm{z}],
\end{align}
where $\bm{h}_k$ is the $k$-th column of $\bm{H}$ and $[\cdot]_{\setminus k}$ denotes that the $k$-th element of the vector is set as zero.

\subsection{Asymptotic Analysis of LMMSE Detector}

According to Eq.~(\ref{AWGN}), a Gaussian assumption is employed to simplify the asymptotic analysis, which is commonly used in~\cite{Lei2016,Lei20162}.

\emph{Assumption 1: The output estimated signal of the LMMSE detector is equivalent to an observation from AWGN channel, i.e., $\bm{\hat{X}}_e=\bm{X}+\hat{\bm{Z}}$, where $\bm{\hat{X}}_e=[\hat{x}_1^e, ..., \hat{x}_K^e]^T, \hat{\bm{Z}}=[\hat{z}_1, ..., \hat{z}_K]^T$, and $\hat{\bm{Z}}$ is an equivalent Gaussian noise with mean $\bm{0}=[0, ..., 0]^T$ and variance $\bm{v}^e=[v_1^e, ..., v_K^e]^T$.
}

With Assumption 1, the output signal of the LMMSE detector can be estimated by tracing the variance of equivalent Gaussian noise $\hat{\bm{Z}}$. That is, when \emph{extrinsic} variance $v_k^e$ decreases to $0$ gradually, $k=1, ..., K$, the estimated signals become more accurate. Note that the update of $\hat{z}_k$ in Eq.~(\ref{AWGN}) is determined by \emph{a priori} variance $v_k$, \emph{a posterior} variance ${\hat{v}_k}$, and \emph{extrinsic} variance $v^e_k$. Therefore, we need to trace the variance updates for the input-output signals of LMMSE detector in the iterative detection process.

Based on \emph{a priori} variance $v_k$, \emph{a posteriori} variance $\hat{v}_{k}$ of signal $\hat{x}_{k}$ from the LMMSE detector~\cite{Lei2016} is calculated by
\begin{small}
\begin{align} \nonumber
&\hat{v}_{k}= v_k\big(1-\frac{1}{4}\mathcal{F}(\frac{K}{M},\frac{\sigma_n^2}{Kv_k})\big),\\ \nonumber
&=\tiny{\frac{\sqrt{({{\sigma_n^2}}/{v_k}+M-K)^2+4K{{\sigma_n^2}}/{v_k}}-({{\sigma_n^2}}/{v_k}+M-K)}{2K(v_k)^{-1}}},
\end{align}
\end{small}
where $\mathcal{F}(a, b)$$=$$\big(\sqrt{(1+1/\sqrt{a})^2+b}-\sqrt{(1-1/\sqrt{a})^2+b}\big)^2$. Then, \emph{extrinsic} variance $v^{e}_k$ of signal $\hat{x}^{e}_k$ is obtained as
\begin{small}
\begin{align} \nonumber
&v^{e}_k=[({\hat{v}_{k}})^{-1}-(v_k)^{-1}]^{-1} \\ \nonumber
&=(v_k){{\frac{\sqrt{({v_k}/{\sigma_n^2}+M-K)^2+4{v_k}/{\sigma_n^2}}-({v_k}/{\sigma_n^2}+M-K)}{({v_k}/{\sigma_n^2}+M+K)-
\sqrt{({v_k}/{\sigma_n^2}+M-K)^2+4K{v_k}/{\sigma_n^2}}}}} \\ \label{LMSE}
&=\frac{\sigma_n^2+cv_k+\sqrt{(\sigma_n^2+cv_k)^2+4\sigma_n^2v_k}}{2},
\end{align}
\end{small}
where $c=\frac{K-M}{M}$. Furthermore, for large-scale systems, i.e., $K\rightarrow \infty$, $M\rightarrow \infty$, and fixed system load $\beta=\frac{K}{M}$, the asymptotic \emph{extrinsic} variance is
\begin{equation}\label{sum_gau}
v^{e}_k=
\left\{ \!\!\!\!\!\!\!{\begin{array}{*{20}{l}}
\begin{array}{l}
~\frac{\sigma_n^2}{M-K}, \qquad~~~~~\;\;\beta < 1,
\end{array}\vspace{3mm}
\\
\begin{array}{l}
~\frac{K-M}{M} v_k, \qquad~~\;\;\;\beta>1,
\end{array}\vspace{1mm}
\\
\begin{array}{l}
~\frac{v_k}{\sqrt{\frac{v_k}{{\sigma_n^2}}K}-1}, \quad~~~~~\;\beta=1.
\end{array}
\end{array}} \right.
\end{equation}

Note that the variance update for estimated signals of LMMSE detector is calculated analytically by using Eq.~(\ref{LMSE}) or Eq.~(\ref{sum_gau}), which makes the asymptotic analysis  easy.

\subsection{Asymptotic Analysis of Decoders}

Although the asymptotic performance analysis of a channel decoder is usually given by the standard EXIT method with a mutual information measure as in~\cite{FangTWC,FangTcom,EXIT,EXFUN,MSE}, to match with our variance transfer analysis of LMMSE detector, we need to transform the mutual information measure into variance measure.

\subsubsection{LMMSE $\rightarrow$ DEC}

Based on Assumption~1, the output estimated signal of LMMSE detector is equivalent to an observation from AWGN channel,
i.e., $\hat{{x}}^e_k=x_k+\hat{{z}}_k$, so that input LLR $L^a_{\rm{DEC}}(x_k)$ of decoder~$k$ associated with $x_k$ is calculated by
\begin{small}
\begin{align} \nonumber
L^a_{\rm{DEC}}(x_k)={\rm{log}}[\frac{P(\hat{{x}}^e_k|{x_k}=+1)}{P(\hat{{x}}^e_k|{x_k}=-1)}]
={\rm{log}}[\frac{{\rm{exp}}(-\frac{(\hat{{x}}^e_k-1)^2}{2v^e_k})}{{\rm{exp}}(-\frac{(\hat{{x}}^e_k+1)^2}{2v^e_k})}]=\frac{2\hat{{x}}^e_k}{v^e_k},
\end{align}
\end{small}
where ${\rm{exp}}(\cdot)$ is the exponential function and ${\rm{log}}[\cdot]$ is the logarithm function with respect to exponential. Then, the mean of $L^a_{\rm{DEC}}(x_k)$ is $m_{k}^a=E[x_kL^a_{\rm{DEC}}(x_k)]=\frac{{2}}{v^e_k}$. According to the Gaussian assumption in EXIT analysis~\cite{EXIT,EXFUN}, $L^a_{\rm{DEC}}(x_k)$ obeys Gaussian distribution $\mathcal{N}(m_k^a, 2m_k^a)$. Thus, \emph{a priori} mutual information $I^{a}_k$ for decoder~$k$ can be calculated by
\begin{equation}\label{Ia}
I^{a}_k=J(\sqrt{2m_k^a})=J(\sqrt{\frac{{4}}{v^e_k}}),
\end{equation}
where $J(\sigma_a)=1-\int_{-\infty}^{+\infty }{\frac{{\rm{exp}}(-\frac{(x-\sigma_a^2/2)^2}{2\sigma_a^2}){\rm{log}}(1+{\rm{exp}}(-x))}{\sqrt{2\pi}\sigma_a}}dx
$.

\subsubsection{EXIT Function of DEC}

Based on the \emph{a priori} mutual information, output mutual information $I_{k}^{e}$ is calculated by the EXIT function of decoder~$k$ and is fed back to the LMMSE detector. Considering that EXIT functions of low-density parity-check (LDPC) like codes are simple~\cite{FangTWC,FangTcom}, the proposed multi-user code is designed based on the structures of LDPC-like codes, such that the corresponding EXIT function can be obtained readily.

\subsubsection{DEC $\rightarrow$ LMMSE}
When the single-user decoding is finished, LLR $L^e_{\rm{DEC}}(\tilde{x}_k)$ associated with output signal $\tilde{x}_k$ of decoder~$k$ is obtained. Then,  the mean and variance of $\tilde{x}_k$ can be calculated by
\begin{align} \nonumber
&E[\tilde{x}_k]={\rm{tanh}}(\frac{L^e_{\rm{DEC}}(\tilde{x}_k)}{2}), \\ \nonumber &Var[\tilde{x}_k]=E[|\tilde{x}_k-x_k|^2|L^e_{\rm{DEC}}(\tilde{x}_k)]=1-(E[\tilde{x}_k])^2.
\end{align}
According to the Gaussian approximation in EXIT analysis~\cite{EXIT,EXFUN}, $L^e_{\rm{DEC}}(\tilde{x}_k)$ obeys Gaussian distribution $\mathcal{N}(m^e_k, 2m^e_k)$, where $m^e_k=\frac{(J^{-1}(I_k^e))^2}{2}$ and function $J^{-1}(I)$ is the inverse of $J(\sigma_a)$. As a result, the variance of $\tilde{x}_k$ that is fed back to the LMMSE detector is
\begin{align} \nonumber
&v_k=E[(\tilde{x}_k-x_k)^2]=E_{L^e_{\rm{DEC}}(\tilde{x}_k)}[Var[\tilde{x}_k]]\\ \label{DEC}
&=E_{L^e_{\rm{DEC}}(\tilde{x}_k)}[1-({\rm{tanh}}(\frac{L^e_{\rm{DEC}}(\tilde{x}_k)}{2}))^2],
\end{align}
where $E_{L^e_{\rm{DEC}}(\tilde{x}_k)}[\cdot]$ is calculated by the Monte Carlo simulations depending on the Gaussian distribution of  $L^e_{\rm{DEC}}(\tilde{x}_k)$.

\begin{algorithm}[!htp]
\caption{Asymptotic EXIT analysis for the iterative receiver}
\begin{algorithmic}[1]
\State {Input $M$, $K$, ${\sigma^{n}}$, initial iteration index $\ell=0$,~$v_k^{0}=1$, $I_k^{a,0}=I_k^{e,0}=0$.
\State \textbf{Repeat:}  set $\ell \Leftarrow \ell+1$,
\State \qquad Calculate \emph{extrinsic} $v_k^{e, \ell}$ of LMMSE detector by using Eq.~(\ref{LMSE}) (using Eq.~(\ref{sum_gau}) for lagre-scale systems).
\State \qquad Transform variance $v_k^{e, \ell}$ into \emph{a priori} $I_k^{a, \ell}$ for decoder~$k$ according to Eq.~(\ref{Ia}).
\State  \qquad Based on the EXIT function of decoder~$k$, update mutual information and then output \emph{extrinsic} $I_k^{e, \ell}$.
\State \qquad Transform $I_k^{e, \ell}$ into \emph{a priori} $v_k^{\ell+1}$ for  LMMSE detector by using Eq.~(\ref{DEC}).
\State \textbf{Until:} $v_k^{e,\ell}=0$ or $I_k^{a,\ell}=I_k^{e,\ell}=1$ (iteration is converged).}
\end{algorithmic}
\end{algorithm}
The complete asymptotic EXIT analysis for the iterative receiver is provided in Algorithm~1. Note that the statistically iterative detection between the LMMSE detector and the decoders is estimated by tracing the variances of estimated signals, which is easy to implement. Meanwhile, by exploiting the proposed asymptotic analysis, the multi-user code based on the structures of LDPC-like codes can be designed and optimized readily.

\section{Practical Coding Scheme for MIMO-NOMA}

In this section, we present a practical coding scheme for MIMO-NOMA system. Subsequently, we analyze the complexity and the asymptotic performance of the overall MIMO-NOMA system.

\subsection{Coding Scheme and Message-Passing Decoder}

\begin{figure}[!htp]
  \centering
  \includegraphics[width=0.8\columnwidth]{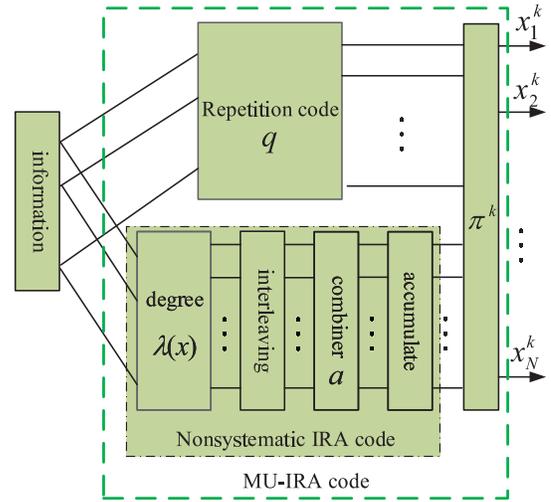}
  \caption{Graph for MU-IRA code structure of user $k$.}\label{FG}
\end{figure}
Since the transmitted signals will be deteriorated by the channel noise and the multi-user interference at the same time, we propose a kind of Multi-User {Irregular Repeat-Accumulate} (MU-IRA) code for the MIMO-NOMA system to overcome both kinds of interferences. Fig.~\ref{FG} shows the graph for MU-IRA code structure of user $k$, which consists of a repetition code, a nonsystematic IRA code~\cite{SLin}, and a user-specific interleaver $\pi^{k}$. The parameters of the MU-IRA code include repetition number $q$, combiner $\alpha$, degree distributions of information sequence $\lambda(x)=\sum_i \lambda_ix^{i-1}$, and code rate $R=\frac{\alpha\sum_i\lambda_i/i}{\alpha q\sum_i\lambda_i/i+1}$.

To explain the advantages of the proposed MU-IRA code, we briefly present the effect of each component in the MU-IRA code.
\begin{itemize}
  \item Although repetition code provides no coding gains in the point-to-point channel, it can provide multi-user coding gains in the multi-user channels to overcome the multi-user interference. For example, spreading in CDMA systems is in fact repetition code, which can achieve coding gains. Previous work~\cite{SongMUCode} theoretically shows that repetition encoding increases the superposed signal distance of multi-user code. Meanwhile, we will show that introducing repetitions in the codeword benefits the iterative processing between the LMMSE detector and channel decoder.
  \item Here, we set combiner $\alpha\geq 1$ in the MU-IRA code, while $\alpha=1$ is in the IRA code designed for Multiple-Access Channel (termed MAC-IRA code) \cite{Song-ISIT2015,Song-TVT2017}. Due to this modification, the proposed MU-IRA code is a generalization of the MAC-IRA code. Note that the multi-user scenarios in \cite{Song-ISIT2015,Song-TVT2017} and this paper are different, where the receiver in \cite{Song-ISIT2015,Song-TVT2017} has a single antenna and the receiver in this paper has multiple antennas. For the single-antenna receiver, each user requires to employ a very low-rate MAC-IRA code with combiner $\alpha=1$ to overcome the severe multi-user interference. In this paper, since the multiple antennas in the receiver can provide power gains to overcome a part of multi-user interference, each user can employ a higher-rate code. Therefore, the proposed MU-IRA code with combiner $\alpha\geq 1$ gives more flexibility to high-rate code design. On the other hand, from the EXIT chart point of view, $\alpha\geq  1$ also provides more flexibility for the decoder's EXIT characteristics, so that we could find better code with EXIT curve matching better with that of the LMMSE detector.
  \item Different interleavers $\pi_k, k=1,...,K$, are employed by different users for user identification~\cite{IDMA}.
\end{itemize}

Now we present the message-passing decoding in the MU-IRA decoder, in preparing for the code parameter optimization. As shown in Fig.~\ref{FG}, the MU-IRA decoder should consist of a repetition decoder, a nonsystematic IRA decoder, and an information combiner. Based on the estimated signals from the LMMSE detector, the repetition decoding and the IRA decoding are performed once parallelly, where the IRA decoding is realized based on the sum product algorithm~\cite{SLin}. Then, the obtained estimations are combined in the information combiner and the generated extrinsic estimations are fed back to the repetition decoder and the IRA decoder according to message update rules \cite{BP-LDPC}. Afterwards, the repetition decoding and the IRA decoding are performed once, where the output estimations are fed back to the LMMSE detector.
\begin{table*}[!htbp]
\caption{Optimized MU-IRA coded MIMO-NOMA systems.}\label{MU-IRA}
\centering
\begin{tabular}{|c|c|c|c|c|c|c|}
\hline
System load & Full loading & Over loading& \multicolumn{4}{c|}{Severe loading} \\
\hline
$\it{\beta}$ & {$1$} & {$2$}& {$3$} & {$4$}&{$8$} & {$8$} \\
\hline
$\textit{K}$ & {$8$} & {$16$}& {$24$} & {$32$} & {$32$} & {$64$}\\
\hline
$\textit{M}$ & {$8$} & {$8$}& {$8$} & {$8$}& {$4$} & {$8$} \\
\hline
$\sigma_n$ & {$4.58$} & {$5.27$}& {$5.52$} & {$6.34$} & {$3.81$} & {$5.43$}\\
\hline
${\textit{R}}$ & $0.2$  & $0.15$  & $0.13$ & $0.1$& {$0.1$} & {$0.1$}\\
\hline
${\textit{R}}_{\text{sum}}$ & $1.6$  & $2.4$  & $3.12$ & $3.2$& {$3.2$} & {$6.4$}\\
\hline
${\textit{q}}$ & $2$  & $2$  & $2$ & $2$& {$4$} & {$4$}\\
\hline
${\it{\alpha}}$ & $4$  & $3$  & $2$ & $2$ & {$2$} & {$2$}\\
\hline
${\it{\lambda}}_{\text{3}}$ & $0.14619$ & $0.129157$ & $0.174135$ & $0.121532$& {$0.207197$} & {$0.204955$}\\
${\it{\lambda}}_{\text{10}}$ & $0.212715$ & $0.173591$ & $0.153139$ & $0.113888$& {$0.036035$} & {$0.044794$}\\
${\it{\lambda}}_{\text{30}}$ &  $0.223699$ & $0.125162$ & $0.254471$ &$0.103885$& {$0.139163$} & {$0.0638$}\\
${\it{\lambda}}_{\text{50}}$ & $0.112159$ &             & $0.085083$ & & {$0.048337$} & {$0.066099$}\\
${\it{\lambda}}_{\text{80}}$ &            & $0.384998$&  $0.333171$ & $0.152555$& {$0.136988$} & {$0.313755$}\\
${\it{\lambda}}_{\text{100}}$ &  $0.305237$ & $0.187092$ &  & $0.50814$ & {$0.43228$} & {$0.306596$}\\
\hline
$(\frac{\textit{E}_{\textit{b}}}{\textit{N}_{\text{0}}})^{\it{\ast}}_{\text{dB}}$ & $-9.22$ & $-9.2$ & $-8.99$ & $-9.05$ & $-4.65$ & $-7.71$ \\
\hline
{MIMO-NOMA}&  \multirow{2}{*}{$-9.39$} &  \multirow{2}{*}{$-9.28$}  &  \multirow{2}{*}{$-9.06$}  &  \multirow{2}{*}{$-9.1$} &  \multirow{2}{*}{$-4.74$}  &  \multirow{2}{*}{$-7.78$}\\
{capacity (dB)} & & & & & &\\
\hline
${\rm{Gap~(dB)}}$ & $0.17$  & $0.08$  & $0.07$ & $0.05$ & $0.09$ & $0.07$\\
\hline
\end{tabular}
\end{table*}
\begin{table*}[!htp]
\centering
\caption{Complexities of the proposed MIMO-NOMA system and each component.}\label{Complexity}
\begin{tabular}{|c|c|c|c|c|}
\hline
\multicolumn{2}{|c|}{Name}                                                                                                                                       & \multicolumn{3}{c|}{Complexity} \\ \hline
\multicolumn{2}{|c|}{$K$-user transmitter}  & \multicolumn{3}{c|}{$\mathcal{O}(K)$}           \\ \hline
\multirow{4}{*}{\begin{tabular}[c]{@{}l@{}}Iterative \\ receiver\end{tabular}} & \multicolumn{1}{c|}{\begin{tabular}[c]{@{}c@{}}LMMSE \\ detector\end{tabular}} & \multicolumn{3}{c|}{$\mathcal{O}((min\{MK^2+K^3, KM^2+M^3\})\tau_{\rm{max}})$}           \\ \cline{2-5}
& \multirow{3}{*}{\begin{tabular}[c]{@{}c@{}}MU-IRA \\ decoders\end{tabular}}     & \multicolumn{3}{c|}{$\mathcal{O}(K\tau_{\rm{max}})$}  \\ \cline{3-5}
&  & $+/-$ & $\times/\div$ &   exp/log  \\ \cline{3-5}
&  &$(\frac{2}{R}+(\frac{1}{R}-q)(5+6\alpha)-1)K\tau_{\rm{max}}$& $(6\alpha(\frac{1}{R}-q))K\tau_{\rm{max}}$& $(6\alpha(\frac{1}{R}-q))K\tau_{\rm{max}}$\\ \hline
\multicolumn{2}{|c|}{Proposed system}                                                                                                                            & \multicolumn{3}{c|}{$\mathcal{O}(min\{MK^2+K^3, KM^2+M^3\}+K)\tau_{\rm{max}}+K)$}           \\ \hline
\end{tabular}
\end{table*}
\subsection{Optimization of Coding Scheme}

\begin{figure}[!htp]
  \centering
  \includegraphics[width=1\columnwidth]{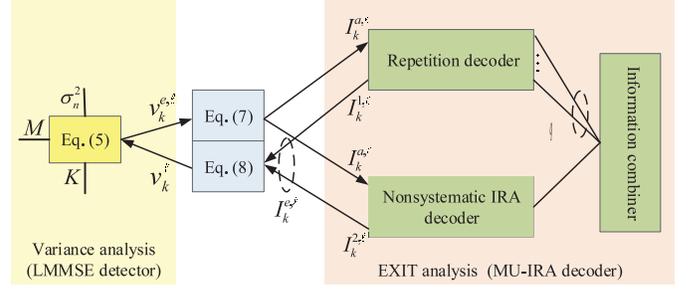}
  \caption{Asymptotic EXIT analyses of the LMMSE detector and the MU-IRA decoder.}\label{VarEXIT}
\end{figure}
With the goal of maximizing sum rate $R_{\rm{sum}}=KR$, the asymptotic EXIT analysis of the iterative receiver is employed to optimize the code parameters whose EXIT property  matches with that of the LMMSE detector. Fig.~\ref{VarEXIT} shows the asymptotic EXIT process between the LMMSE detector and the MU-IRA decoder, where the EXIT function of the MU-IRA decoder is obtained by using the similar method in~\cite{Song-TVT2017}. According to Algorithm 1, \emph{a priori} variance $v^{\ell}_{k}$ and \emph{extrinsic} variance $v^{e,\ell}_k$ of LMMSE detector are updated according to Eq.~(\ref{LMSE}), Eq.~(\ref{Ia}), and Eq.~(\ref{DEC}). Based on \emph{a priori} mutual information $I_k^{a, \ell}$ and the message-passing decoding, the EXIT function of the MU-IRA decoder is performed to obtain \emph{extrinsic} mutual information $(I_k^{1, \ell}, I_k^{2, \ell})$, which are combined as \emph{extrinsic} $I_k^{e,\ell}$ feeding back to the LMMSE detector. Due to the fact that there are multiple optimized parameters for different repetition number $q$, we should choose the optimal parameters. To be specific, let the maximum repetition number be $q_{\rm{max}}$. For $q=1, ..., q_{\rm{max}}$, based on the asymptotic EXIT analysis, the optimized code is obtained by optimizing $\lambda(x)$ and $\alpha$. Among these candidate codes, the optimal code with the maximum sum rate is selected.

For example, we optimize MU-IRA codes over MIMO-NOMA systems with three types of system loads, i.e., full loading (${\it{\beta}}=1, K=8, M=8$), over loading (${\it{\beta}}=2, K=16, M=8$), and severe system loading (${\it{\beta}}=3, K=24, M=8$), (${\it{\beta}}=4, K=32, M=8$), (${\it{\beta}}=8, K=32, M=4$), and (${\it{\beta}}=8, K=64, M=8$). Note that in MIMO-NOMA scenarios, overloading and severe loading denote that the number of transmitted data streams is larger than the spatial degrees of freedom, i.e., the product of user number and the number of transmit antennas is larger than the number of receive antennas. In this case, the system multiplexing gain is maximal. The corresponding noise variance is $\sigma_n \in \{4.58$, $5.27$, $5.52$, $6.34$, $3.81$, $5.43\}$ and $q_{\rm{max}}=5$. The optimized code parameters are presented in Table~\ref{MU-IRA}, which shows that the decoding thresholds $(\frac{\textit{E}_{\textit{b}}}{\textit{N}_{\text{0}}})^{\it{\ast}}_{\text{dB}}$ of MU-IRA coded MIMO-NOMA systems are within $0.2$~dB from the corresponding system capacities.

We also observed that the optimal value of $q$ could increase to 4 as the system load increases to 8. This is because when the multiplexing gain is large, more repetitions are needed to deal with the signal interference.
\subsection{Complexity Analysis}

To verify practicability of the proposed system, we investigate the implementation complexity for the overall system. In the $K$-user transmitters, since the calculations of encoding and modulation are just additions and modulo operations, the implementation complexity is $\mathcal{O}(K)$. In the iterative receiver, the complexity of LMMSE detector is $\mathcal{O}(min\{MK^2+K^3,KM^2+M^3\}\tau_{\rm{max}})$, where $\tau_{\rm{max}}$ is the maximum iteration number. In the MU-IRA decoder, the averaged number of addition/subtraction ($+/-$), multiplication/division ($\times/\div$), and exponent/logarithm (exp/log) operations is $(\frac{2}{R}+(\frac{1}{R}-q)(5+6\alpha)-1)$, $6\alpha(\frac{1}{R}-q)$, and $6\alpha(\frac{1}{R}-q)$ for decoding one information bit of MU-IRA code in one iteration. As a result, the decoding complexity of $K$ information bits in $K$-user decoders is approximately $\mathcal{O}(K\tau_{{\rm{max}}})$. Note that the implementation complexity of $K$-user decoders decreases with the increase of $q$. In summary, the complexities of the overall system and each component are given in Table~\ref{Complexity}.
\begin{figure*}[!htbp]
\centering
\subfigure[Full loading ($K=8, M=8$): decoding trajectory between LMMSE detector and the rate-$0.2$ proposed MU-IRA code.]{
  \includegraphics[width=8cm,height=7cm]{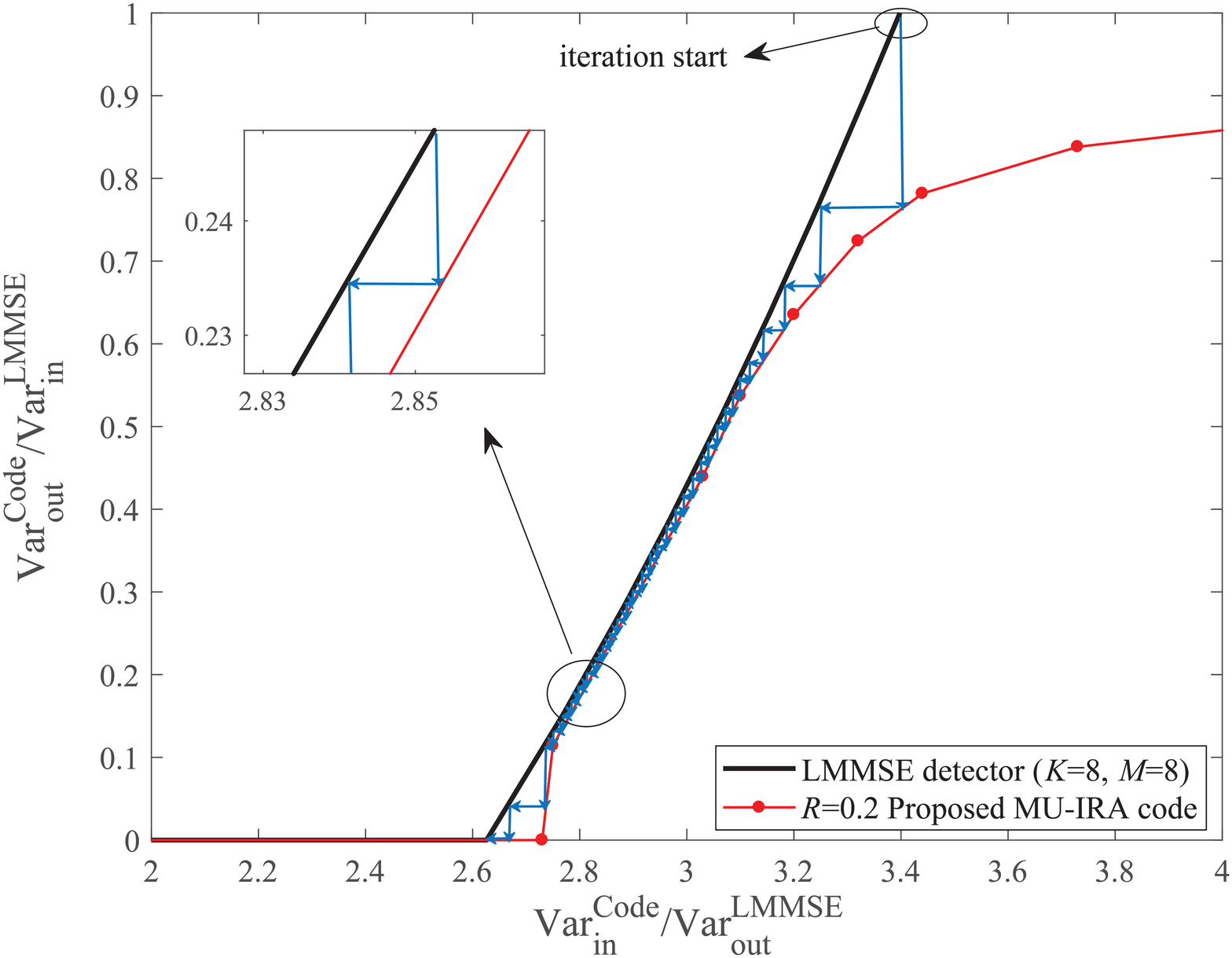}}
\hspace{2ex}
\subfigure[Full loading $(K=8, M=8)$: advantage analysis for the rate-$0.2$ proposed MU-IRA code.]{%
  \includegraphics[width=8cm,height=7cm]{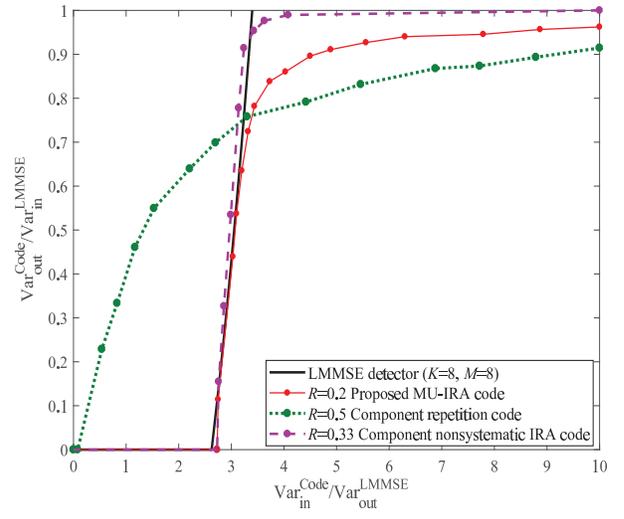}}\\
  \subfigure[Full loading $(K=8, M=8)$: decoding failure region of the rate-$0.2$ SU-IRA code and rate loss of the rate-$0.08$ MAC-IRA code~\cite{Song-ISIT2015,Song-TVT2017} from the rate-$0.2$ proposed MU-IRA code.]{%
  \includegraphics[width=8cm,height=7cm]{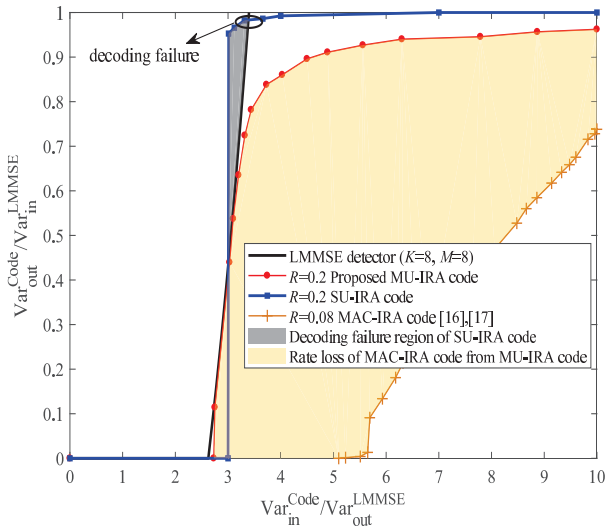}}
\hspace{2ex}
\subfigure[Severe loading $(K=32, M=8)$: decoding failure region of the rate-$0.1$ SU-IRA code and rate loss of the rate-$0.08$ MAC-IRA code~\cite{Song-ISIT2015,Song-TVT2017} from the rate-$0.1$ proposed MU-IRA code.]{%
  \includegraphics[width=8cm,height=7cm]{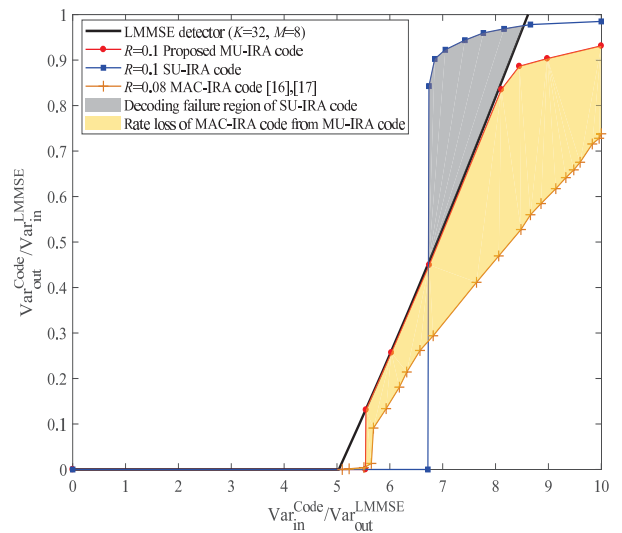}}
\caption{Variance-transfer curves of the LMMSE detector over full loading and severe loading MIMO-NOMA, i.e., $(K, M)=(8, 8)$ and $(32, 8)$, the rate-$0.2$ and rate-$0.1$ MU-IRA codes in Table~\ref{MU-IRA}, the component repetition code and nonsystematic IRA code of the rate-$0.2$ MU-IRA code, the rate-$0.08$ MAC-IRA code~\cite{Song-ISIT2015,Song-TVT2017}, and the rate-$0.2$ and rate-$0.1$ SU-IRA codes in Tabel~\ref{SU-IRA}.}
\label{varTran}
\end{figure*}
\subsection{Asymptotic Performance Analysis}

To illustrate that the proposed system can achieve capacity-approaching performances, we provide the input-output variance-transfer curves of the LMMSE detector and those of the MU-IRA codes in Table~\ref{MU-IRA} over full loading and severe loading MIMO-NOMA, i.e., ($K, M$) $=$ $(8,8)$ and ($32, 8$).

As shown in Fig.~\ref{varTran}(a) and Fig.~\ref{varTran}(d), the variance-transfer curves of rate-$0.2$ and rate-$0.1$ MU-IRA codes match well with those of the LMMSE detector over full loading and severe loading cases. According to the capacity-achieving proof of the LMMSE detection in~\cite{Lei-ICC}, the proposed MU-IRA coded system can approach the capacity of MIMO-NOMA system. In addition, Fig.~\ref{varTran}(a) also shows the iterative decoding trajectory between the LMMSE detector and the MU-IRA decoder.

To see the advantage of the MU-IRA code in the iterative decoding process, Fig.~\ref{varTran}(b) shows the variance-transfer curves for component codes of the rate-$0.2$ MU-IRA code in Table~\ref{MU-IRA}, i.e., a rate-$0.5$ repetition code and a rate-$0.33$ nonsystematic IRA code. Notice that the repetition code can provide the high output SNR (i.e., $1/{\rm{Var}}_{\rm{out}}^{\rm{Code}}$) when the input SNR (i.e., $1/{\rm{Var}}_{\rm{in}}^{\rm{Code}}$) is relatively low. By contrast, the nonsystematic IRA code can provide the very high output SNR when the input SNR is medium or large. As a result, the proposed MU-IRA code combines the advantages of these two component codes, which aid the variance-transfer curve of the MU-IRA code to match well with that of the LMMSE detector in the entire SNR region.

To confirm the importance of matching between the LMMSE detector and the proposed code in the perspective of EXIT analysis, we present a rate-$0.08$ MAC-IRA code~\cite{Song-ISIT2015,Song-TVT2017} for comparison, whose parameters are $\lambda(x)=0.063021x+0.228288x^2+0.111951x^9+0.226877x^{29}+0.369864x^{49}$, $\alpha=1$, and $q=5$. As shown in Fig.~\ref{varTran}(c) and Fig.~\ref{varTran}(d), the variance-transfer curves of MAC-IRA code seriously mismatch with those of the LMMSE detector over the full loading and severe loading cases, in which the large gaps between variance-transfer curves of the MAC-IRA code and those of the proposed MU-IRA codes denote large rate losses.

To emphasize the necessity of multi-user code design, we consider conventional Single-User (SU) IRA codes~\cite{SLin} for comparison. To compare with our code, We design a SU-IRA code for the point-to-point channel by using the EXIT analysis. The optimized parameters of the SU-IRA codes are given in Table~\ref{SU-IRA}, which shows that the decoding thresholds are about $0.1$~dB from the capacity of the point-to-point channel. However, when we use the codes in MIMO-NOMA system, their the variance-transfer curves untimely interact with those of the LMMSE detector as shown in Fig.~\ref{varTran}(b) and Fig.~\ref{varTran}(c), which result in decoding failures.
Moreover, as the system load increases, the decoding failure regions of the SU-IRA codes become large. This indicates that the well-designed SU-IRA codes do not be suitable for the MIMO-NOMA system with the LMMSE detector.
\vspace{-0.1cm}
\begin{table}[!tbp]
\caption{Well-designed SU-IRA codes}\label{SU-IRA}
\centering
\begin{tabular}{|c|c|c|c|c|}
\hline
${\textit{R}}$ & $0.2$  & $0.15$  & $0.13$ & $0.1$\\
\hline
${\textit{q}}$ & $1$  & $1$  & $1$ & $1$\\
\hline
${\it{\alpha}}$ & $4$  & $3$  & $2$ & $2$ \\
\hline
${\it{\lambda}}_{\text{3}}$ & $0.099822$ & $0.091575$ & $0.118814$ & $0.085867$\\
${\it{\lambda}}_{\text{10}}$ & $0.214201$ & $0.171829$ & $0.204525$ & $0.132226$\\
${\it{\lambda}}_{\text{30}}$ &  $0.023108$ & $0.122928$ & $0.196695$ &$0.198883$\\
${\it{\lambda}}_{\text{50}}$ &              &             & $0.346954$ & \\
${\it{\lambda}}_{\text{80}}$ & $0.186412$ & $0.278914$&  $0.016878$ & $0.276011$\\
${\it{\lambda}}_{\text{100}}$ &  $0.476457$ & $0.334754$ & $0.116134$ & $0.307013$ \\
\hline
$(\frac{\textit{E}_{\textit{b}}}{\textit{N}_{\text{0}}})^{\it{\ast}}_{\text{dB}}$ & $-0.8$ & $-1.05$ & $-1.11$ & $-1.24$ \\
\hline
{Point-to-}&  \multirow{3}{*}{$-0.96$} &  \multirow{3}{*}{$-1.13$}  &  \multirow{3}{*}{$-1.19$}  &  \multirow{3}{*}{$-1.29$} \\
{Point channel} & & & &\\
{capacity (dB)} & & & &\\
\hline
\end{tabular}
\end{table}

\section{Numerical Results}

The above analyses and optimizations are based on the assumptions of infinite code length and iterations. To verify the practicability and reliability of the proposed MIMO-NOMA system, in this section, we provide extensive finite-length simulations in various aspects.

\subsection{Comparisons of Decoding Complexity}

The complexity comparisons of the proposed MU-IRA code and the SU-IRA codes are given in Table \ref{Complexity2}, which focuses on the decoding for one information bit of each user per iteration. According to Table~\ref{Complexity}, the number of operations including ${+/-}$, ${\times/\div}$, and exp/log are calculated, where the parameters of the MU-IRA codes and the SU-IRA codes are given in Table~\ref{MU-IRA} and Table~\ref{SU-IRA} respectively. Table~\ref{Complexity2} demonstrates that the proposed MU-IRA codes achieve lower decoding complexities than the SU-IRA codes. As a result, the MU-IRA coded MIMO-NOMA systems have lower implementation complexities than the SU-IRA coded MIMO-NOMA systems.

\begin{table}[!htp]
\caption{Decoding Complexity of one information bit of each user per iteration for the MU-IRA codes in Table~\ref{MU-IRA} and the SU-IRA codes in Table~\ref{SU-IRA}.} \label{Complexity2}
\centering
\begin{tabular}{|c|c|c|c|c|c|}
\hline
System load & Code & Rate& ${+/-}$ & ${\times/\div}$ & exp/log\\
\hline
{Full loading} & MU-IRA & 0.2 & 768 & 576 & 576\\
\cline{2-6}
 {($K=8, M=8$)}& SU-IRA &0.2  &1000& 768&768\\
\hline
{Over loading} & MU-IRA &0.15 &1915& 1344& 1344\\
\cline{2-6}
  {($K=16, M=8$)}& SU-IRA &0.15   &2283& 1632& 1632\\
\hline
{Severe loading} & MU-IRA &0.13  & 2668 & 1639 & 1639\\
\cline{2-6}
  {($K=24, M=8$)}& SU-IRA &0.13   &3076& 1927&1927 \\
\hline
{Severe loading} & MU-IRA &0.1  & 4960 & 3072 & 3072 \\
\cline{2-6}
 {($K=32, M=8$)}& SU-IRA &0.1   & 5504 &3456 & 3456\\
\hline
\end{tabular}
\end{table}

\subsection{Performance Comparison}
To confirm the advantage of the proposed MIMO-NOMA system,  we present the comparisons of three coded  MIMO-NOMA systems, which are the proposed MU-IRA coded system with LMMSE detector, denoted as LMMSE+MU-IRA, the SU-IRA coded system with LMMSE detector, denoted as LMMSE+SU-IRA, and the SU-IRA coded system with the MUD consisting of LMMSE detector and SIC detector~\cite{Tse}, denoted as LMMSE-SIC+SU-IRA.

We consider that the information length of each user is $4096$, the repetition pattern of repetition code is ${\bm{G}}_{\rm{rep}}=[+1,-1,+1,-1,..., +1]$, and a random interleaver is employed in the MU-IRA code. Assume that each user employs BPSK and SNR $E_b/N_0=\frac{P}{2R\sigma_n^2}$, where $P=1$ is the transmitted power of each user. The elements of channel matrix $\bm{H}$ obey a real Gaussian distribution $\mathcal{N}(0, 1)$ and the maximum iteration number is $\tau_{\rm{max}}=250$.

\begin{figure}[!t]
\centering
\subfigure[Full loading ($K=8, M=8$)]{
  \includegraphics[width=8cm,height=7cm]{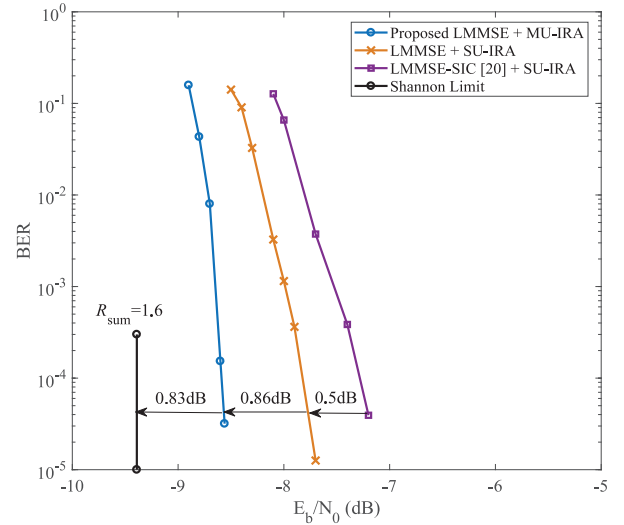}}\hfill
\subfigure[Over loading ($K=16, M=8$)]{%
  \includegraphics[width=8cm,height=7cm]{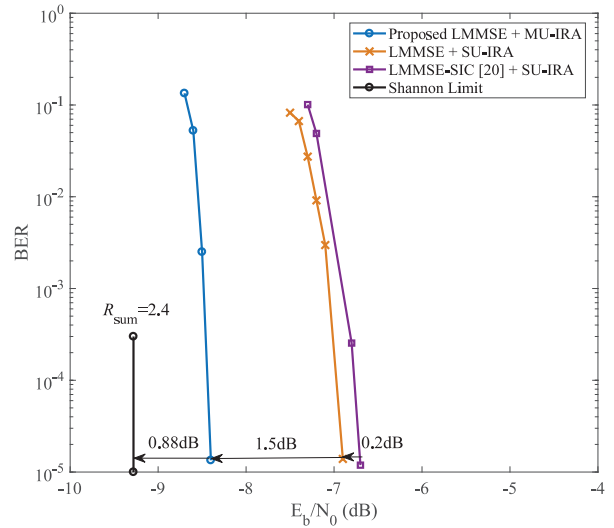}}
\caption{BER curves of three kinds of coded MIMO-NOMA systems with full loading ($K=8, M=8$) and over loading ($K=16, M=8$). These three systems are the proposed MU-IRA coded system with LMMSE detector, the SU-IRA coded system with LMMSE detector, and the SU-IRA coded system with the MUD consisting of LMMSE detector and SIC detector~\cite{Tse}, which are denoted as LMMSE+MU-IRA, LMMSE+SU-IRA, and LMMSE-SIC+SU-IRA, respectively.}
\label{SimMU-IRA}
\end{figure}
Fig.~\ref{SimMU-IRA} provides the bit-error rate (BER) simulations of these three coded systems over the full loading and over loading MIMO-NOMA, where $(K, M)$ $=$ $(8, 8)$ and $(16, 8)$. Note that the gaps between BER curves at $10^{-4}$ of the proposed LMMSE+MU-IRA systems and the corresponding Shannon limits are $0.83$~dB and $0.88$~dB respectively. This verifies that the proposed system can achieve capacity-approaching performances.

Compared with the LMMSE+SU-IRA systems, the proposed systems have $0.86$ dB and $1.5$~dB performance gains in the full loading and over loading cases. This indicates when the system load increases, the proposed system can achieve more performance gains. By comparing two SU-IRA coded systems, Fig.~\ref{SimMU-IRA} shows that the LMMSE+SU-IRA systems can achieve $0.5$ dB and $0.2$ dB performance gains over the LMMSE-SIC+SU-IRA systems, which indicates that the joint iterative multi-user decoding is more reliable than the SIC receiver.

\subsection{The Importance of EXIT Matching between LMMSE detector and Message-Passing decoders }
\begin{figure}[!t]
\centering
\subfigure[Full loading ($K=8, M=8$)]{
  \includegraphics[width=8cm,height=7cm]{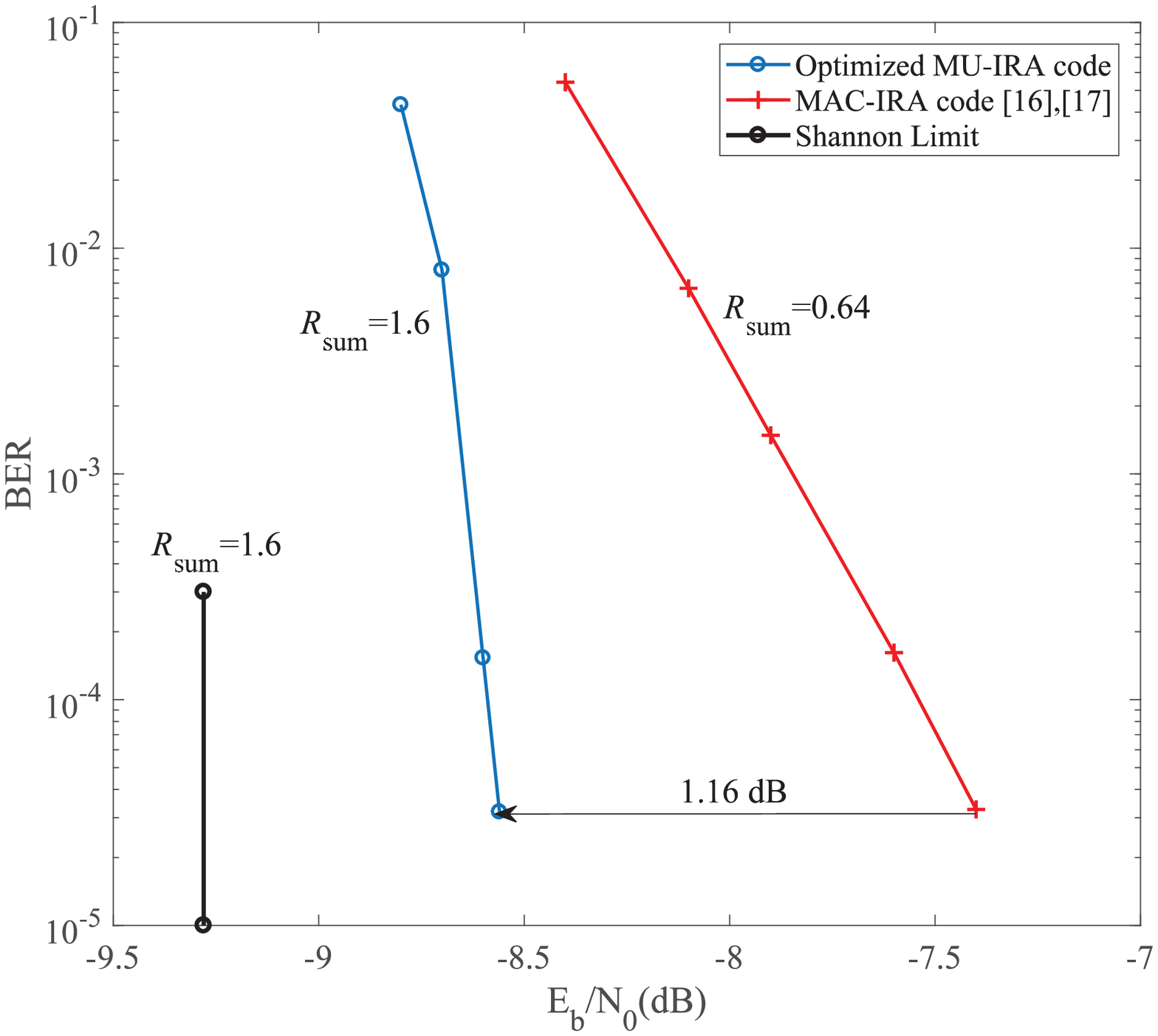}}\hfill
\subfigure[Over loading ($K=16, M=8$)]{%
  \includegraphics[width=8cm,height=7cm]{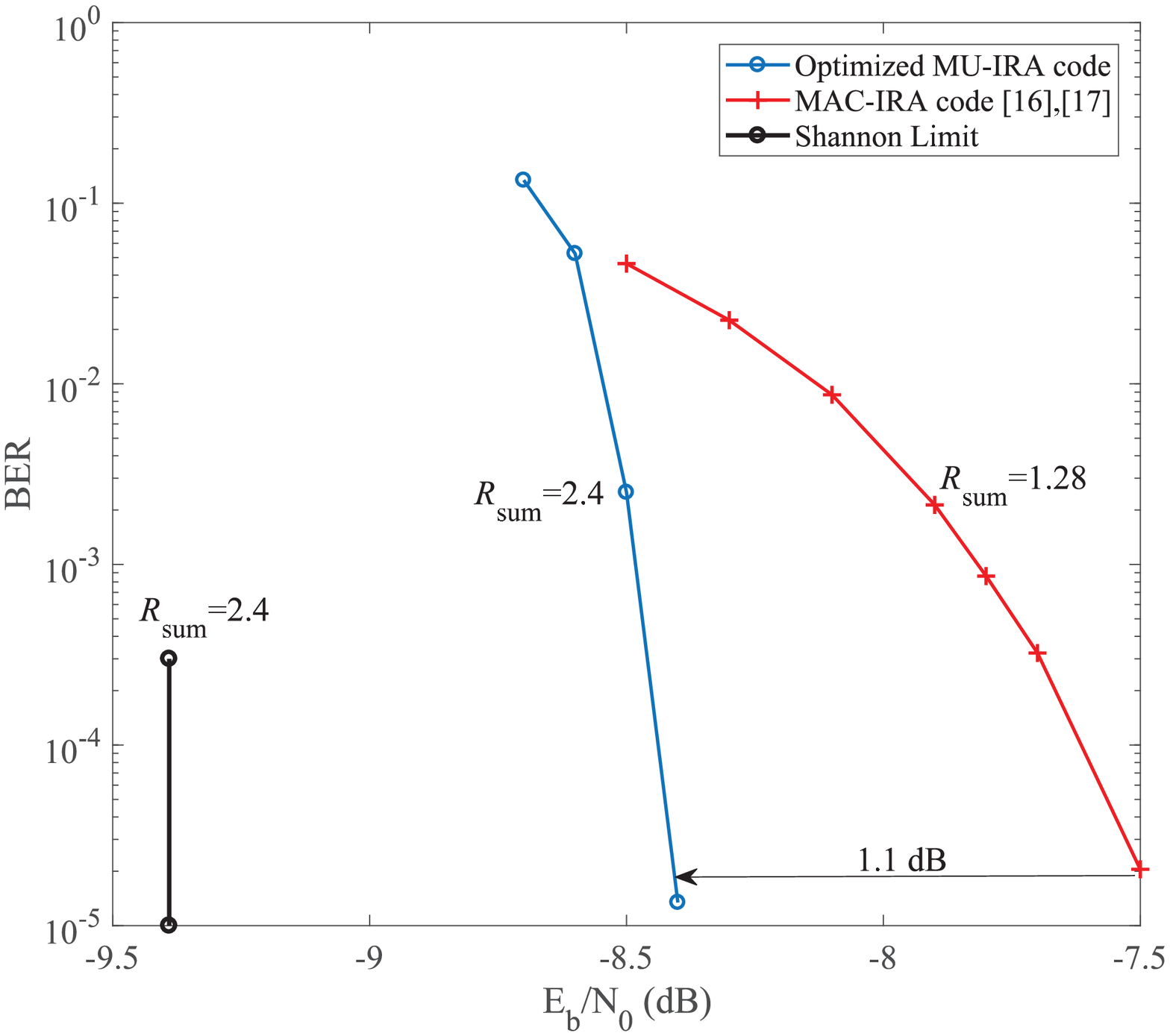}}
\caption{BER curves of the proposed MU-IRA coded systems and the MAC-IRA~\cite{Song-ISIT2015,Song-TVT2017} coded systems over full loading ($K=8, M=8$) and over loading ($K=16, M=8$).}
\label{SimRN-IRA}
\end{figure}
Fig.~\ref{SimRN-IRA} compares the rate-$0.08$ MAC-IRA~\cite{Song-ISIT2015,Song-TVT2017} coded MIMO-NOMA with the rate-$0.2$ and rate-$0.15$ MU-IRA coded MIMO-NOMA. Note that the MU-IRA coded systems have $1.16$~dB and $1.1$~dB performance gains as well as $0.96$ and $1.12$ sum-rate gains over the MAC-IRA coded systems in full loading and over loading cases respectively. This demonstrates the necessity of EXIT matching between the LMMSE detector and the  message-passing decoders.

\subsection{Impact of Code Length}
\begin{figure}[!t]
\centering
\subfigure[Full loading $(K=8, M=8)$]{
  \includegraphics[width=8cm,height=7cm]{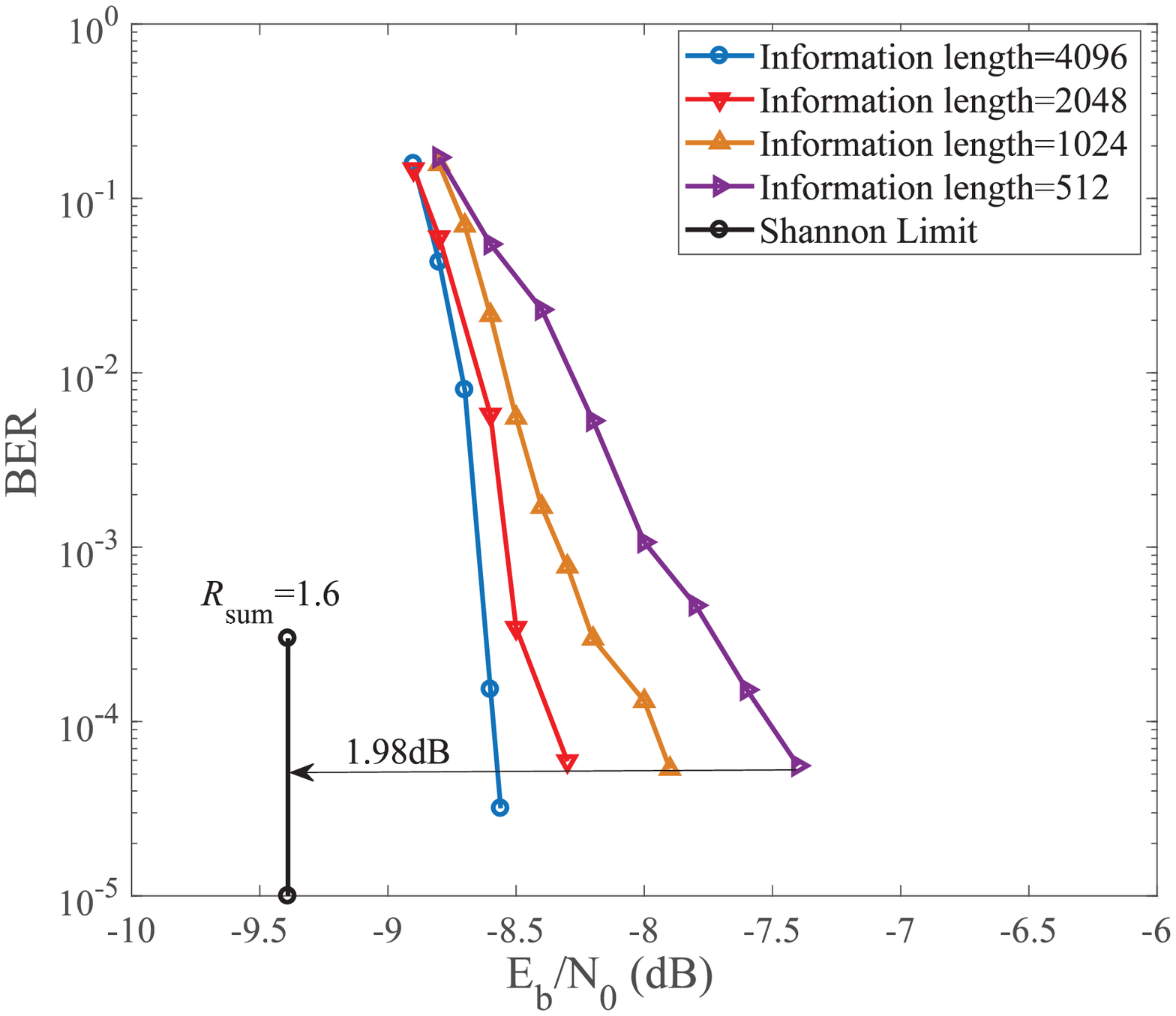}} \hfill
\subfigure[Severe loading $(K=24, M=8)$]{%
  \includegraphics[width=8cm,height=7cm]{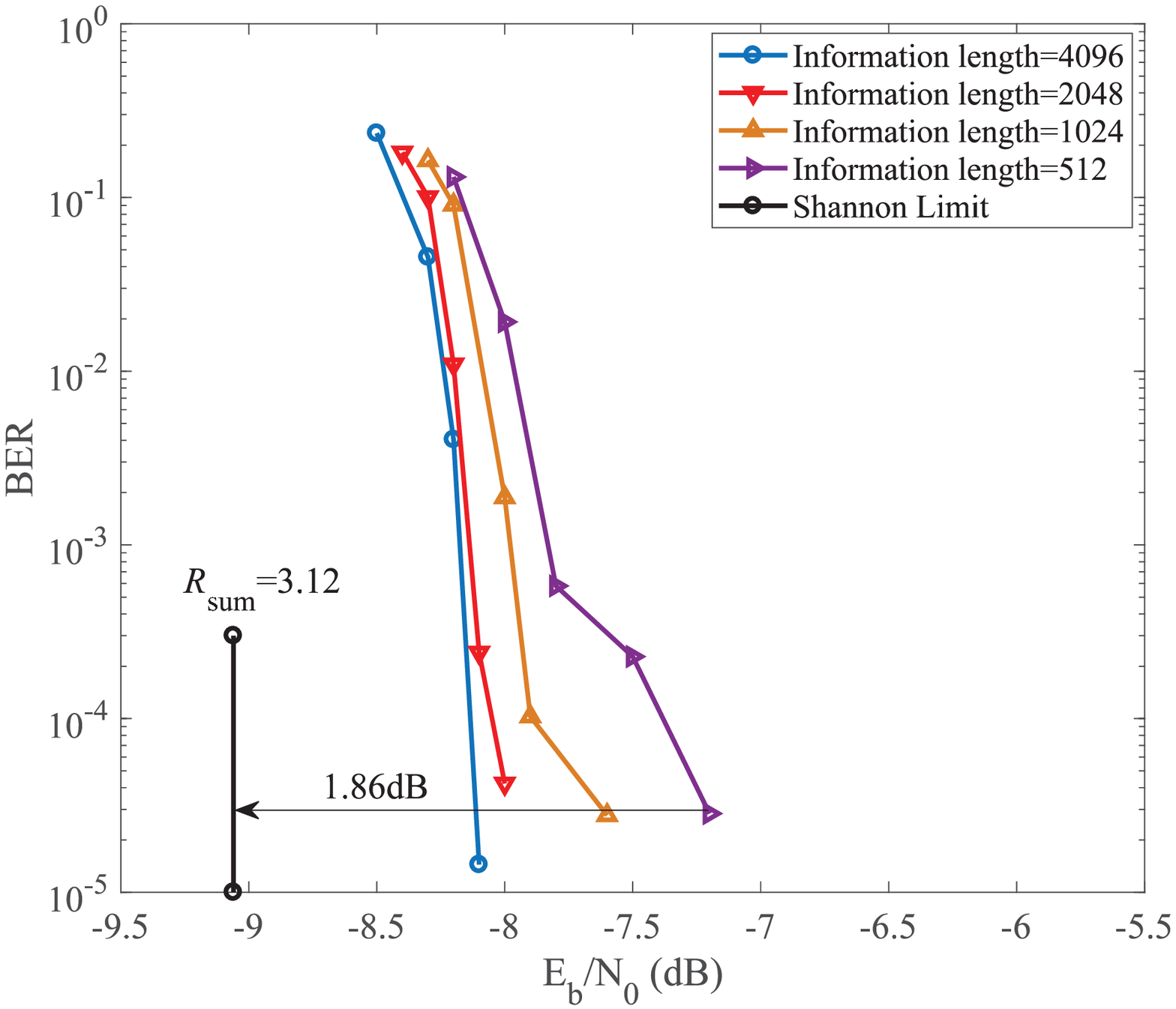}}
\caption{BER curves of the proposed MU-IRA coded MIMO-NOMA systems in full loading ($K=8, M=8$) and severe loading ($K=24, M=8$) cases, wherein the information lengths are $4096, 2048, 1024$, and $512$.}
\label{ShortCode}
\end{figure}

In the practical applications, different code lengths might be required. Hence, we investigate the impact of the finite-length MU-IRA codes on the proposed system, where the information lengths are $4096$, $2048$, $1024$, and $512$. Fig.~\ref{ShortCode} shows the BER performances of the MU-IRA codes obtained in Table~\ref{MU-IRA} over the full loading ($K=8, M=8$) and severe loading ($K=24, M=8$) MIMO-NOMA. Due to the shortened code length, some performance losses are caused. Nevertheless, the gaps between the BER curves at $10^{-4}$ of the MU-IRA code with the shortest information length, i.e., $512$, and the corresponding Shannon limits are still within $2$~dB. This further confirms the practicability of the proposed system.

\subsection{Impact of Iteration Number}
\begin{figure*}[!t]
\centering
\subfigure[Full loading ($K=8, M=8$)]{
  \includegraphics[width=8cm,height=7cm]{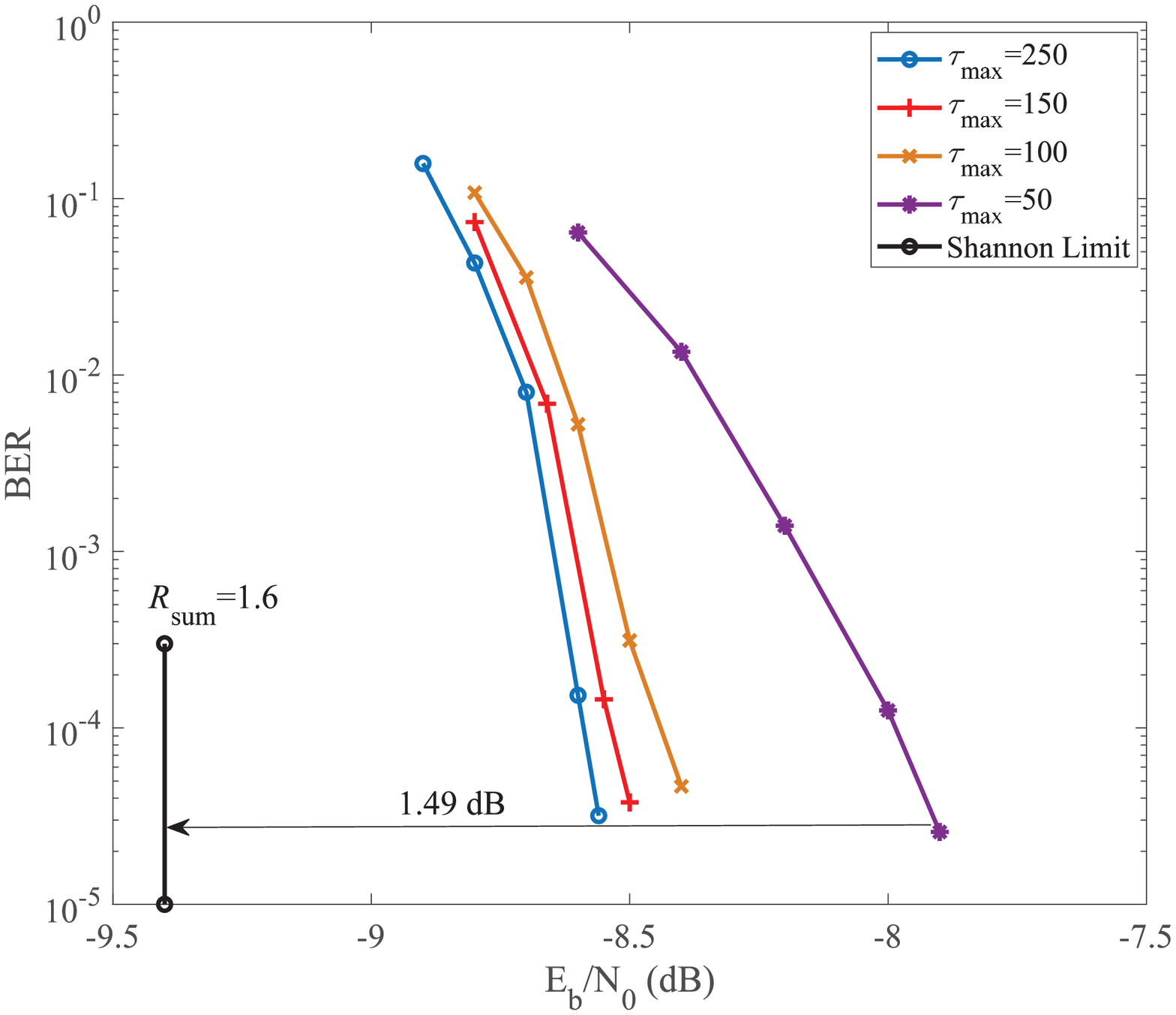}}
  \hspace{2ex}
\subfigure[Over loading ($K=16, M=8$)]{%
  \includegraphics[width=8cm,height=7cm]{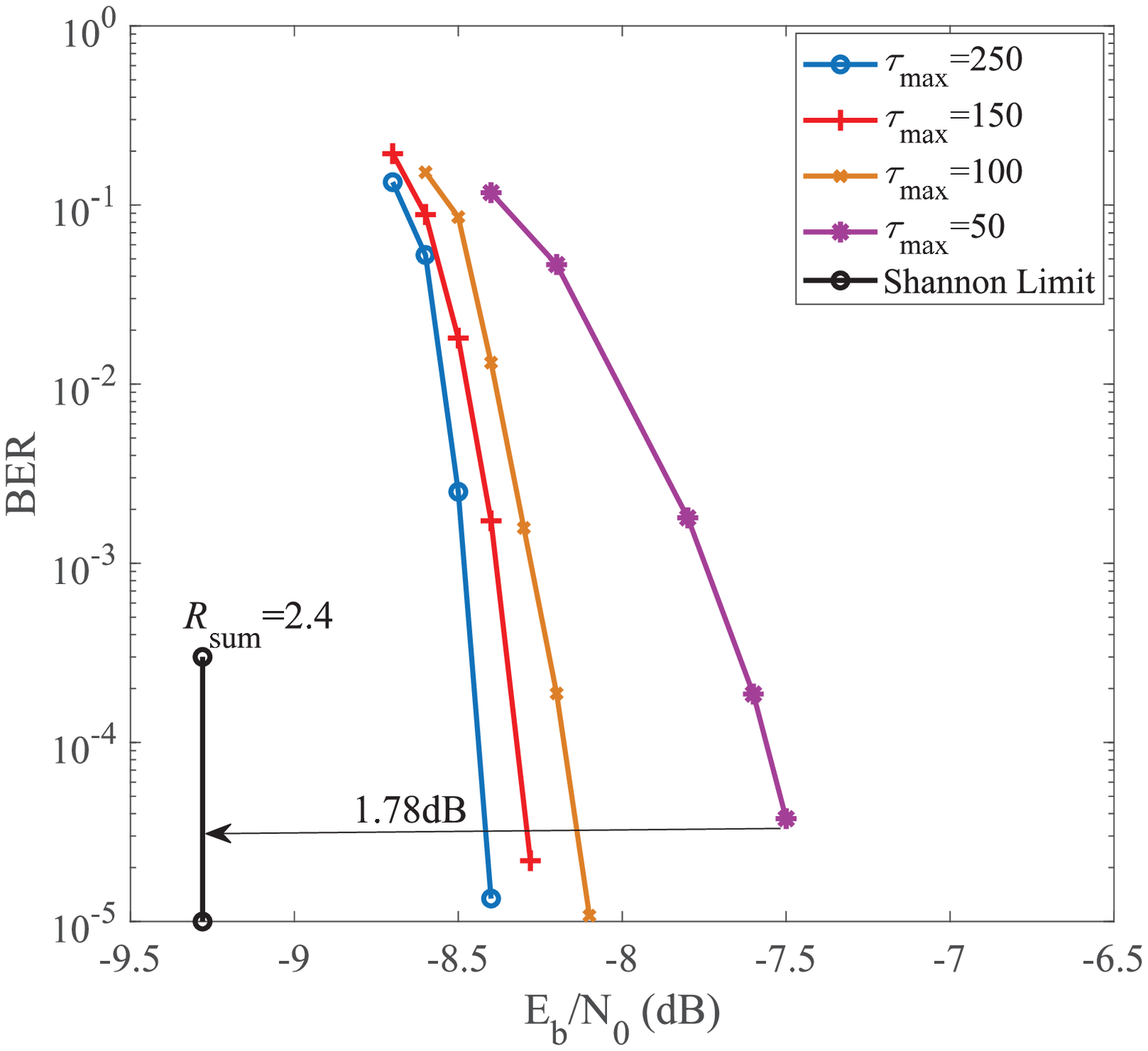}}
\caption{BER curves of the MU-IRA coded systems obtained in Table I under the maximum iteration number $\tau_{\rm{max}}\in\{250, 150, 100, 50\}$ over full loading MIMO-NOMA ($K=8, M=8$) and over loading MIMO-NOMA ($K=16, M=8$).}
\label{Iter¡ªIRA}\vspace{0.1cm}
\end{figure*}

Due to the requirement of low-latency communication, low iteration number should be considered. To investigate the impact of iteration number on the proposed system, Fig.~\ref{Iter¡ªIRA} shows the BER performances of the MU-IRA coded systems with the maximum iteration number $\tau_{\rm{max}}\in\{250, 150, 100, 50\}$ over full loading MIMO-NOMA ($K=8, M=8$) and over loading MIMO-NOMA ($K=16, M=8$). Note that the performance gaps between the MU-IRA codes with $\tau_{\max}=100$ and the MU-IRA codes with $\tau_{\max}=250$ are just $0.3$~dB. When $\tau_{\max}=50$, the gaps between the BER curves at $10^{-4}$ of the MU-IRA codes and the corresponding Shannon limits are within $1.8$ dB. This validates the fast convergence of the proposed codes.

\subsection{Dynamic system load}
\begin{figure*}[!htbp]
\centering
\subfigure[BER curves of the MU-IRA code designed for the full loading case ($K=8, M=8$)($\beta_{\rm{design}}=1$) over changing load cases, i.e., $\beta_{\rm{real}} \in \{0.5, 0.75, 1.25, 1.5\}$.]
{\includegraphics[width=15cm,height=7cm]{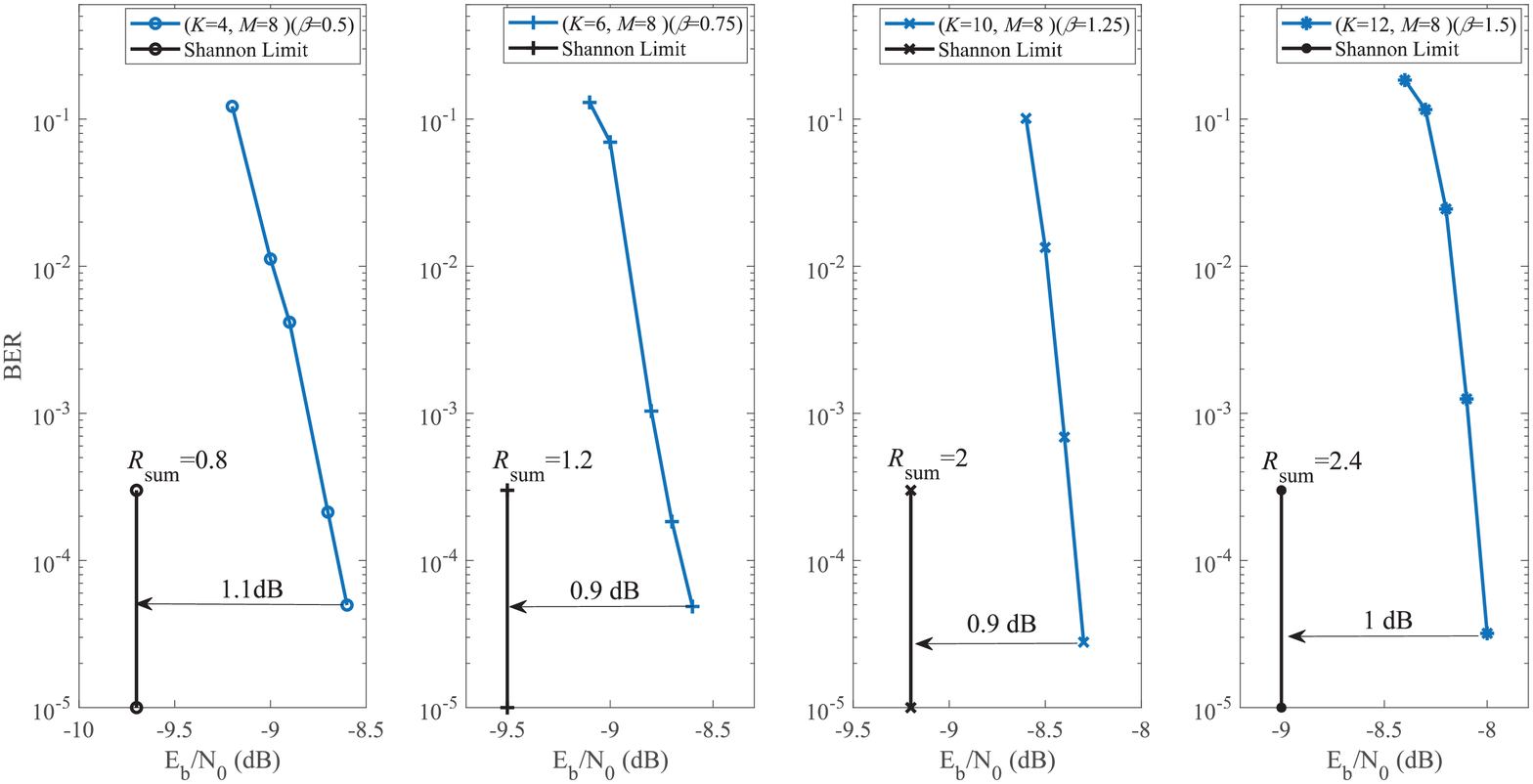}}\vfill
\subfigure[BER curves of the MU-IRA code designed for the over loading case ($K=16, M=8$)($\beta_{\rm{design}}=2$) over changing load cases, i.e., $\beta_{\rm{real}} \in \{1.5, 1.75, 2.25, 2.5\}$.]{%
  \includegraphics[width=15cm,height=7cm]{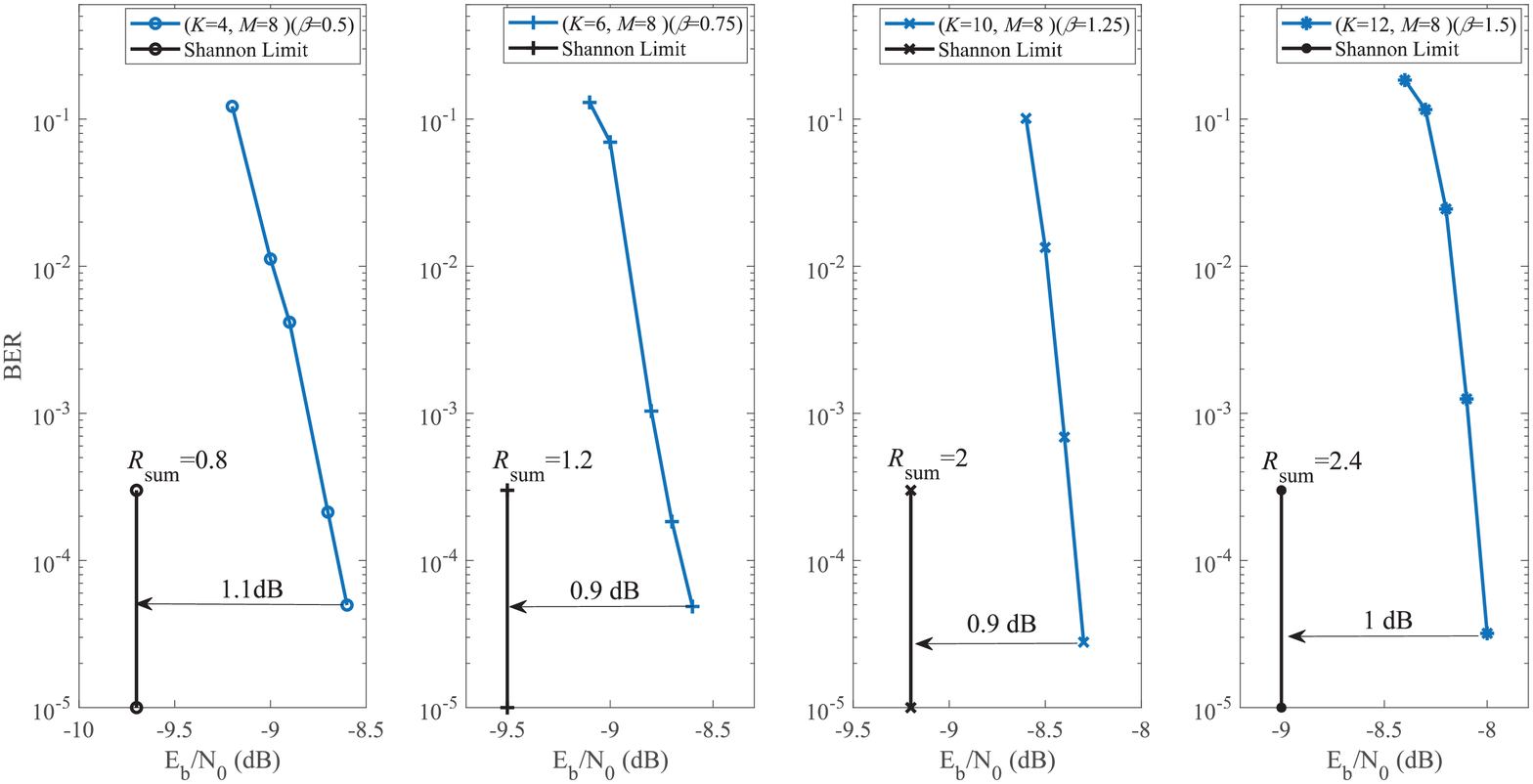}}
\caption{BER curves of the MU-IRA code designed for full loading case ($K=8, M=8$) and over loading case ($K=16, M=8$) over changing load cases $\beta_{\rm{real}} \in \{0.5, 0.75, 1.25, 1.5\}$ and $\beta_{\rm{real}} \in \{1.5, 1.75, 2.25, 2.5\}$ respectively.}
\label{DYRN-IRA}
\end{figure*}
In practice, some users will leave the system when finished communications and some new users will joint the system when ready for communications. As a result, the system load will be dynamic over times. To investigate the robustness of the proposed system over the changing system load cases, Fig.~\ref{DYRN-IRA} shows the BER performances of the MU-IRA codes over the different load cases $\beta_{\rm{real}} \in \{0.5, 0.75, 1.25, 1.5\}$ and $\beta_{\rm{real}} \in \{1.5, 1.75, 2.25, 2.5\}$ respectively, where the MU-IRA code designed for full loading case ($K=8, M=8$)($\beta_{\rm{design}}=1$) is simulated for $\beta_{\rm{real}} \in \{0.5, 0.75, 1.25, 1.5\}$ and the MU-IRA code designed for over loading case ($K=16, M=8$)($\beta_{\rm{design}}=2$) is simulated for  $\beta_{\rm{real}} \in \{1.5, 1.75, 2.25, 2.5\}$. Note that the gaps between BER curves at $10^{-4}$ of the MU-IRA codes and the corresponding capacities are still within 1.45~dB, which illustrates that the proposed system is robust and can provide reliable performances over the low load and changing load cases.

\subsection{Impact of Channel Correlation and Imperfect CSI}

In above simulations, we consider the fast fading channel and the receiver can obtain the perfect CSI. To investigate the robustness of the proposed system, we investigate the impacts of channel correlation and imperfect CSI on the proposed system as follows.

\subsubsection{Block Fading Channel}
\begin{figure*}[!t]
\centering
\subfigure[Full loading $(K=8, M=8)$]{
  \includegraphics[width=8cm,height=7cm]{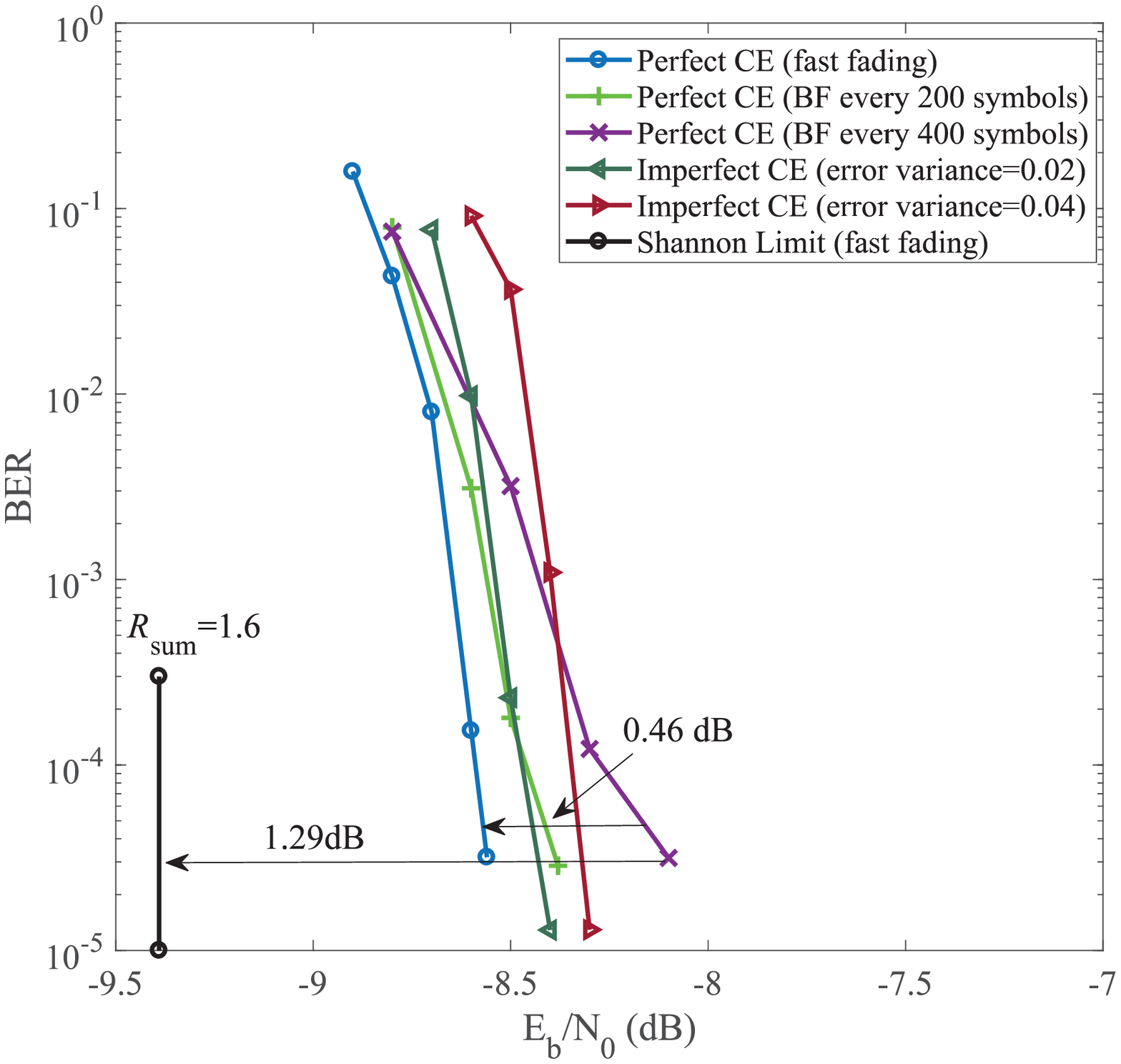}}
  \hspace{2ex}
\subfigure[Severe loading $(K=32, M=8)$]{%
  \includegraphics[width=8cm,height=7cm]{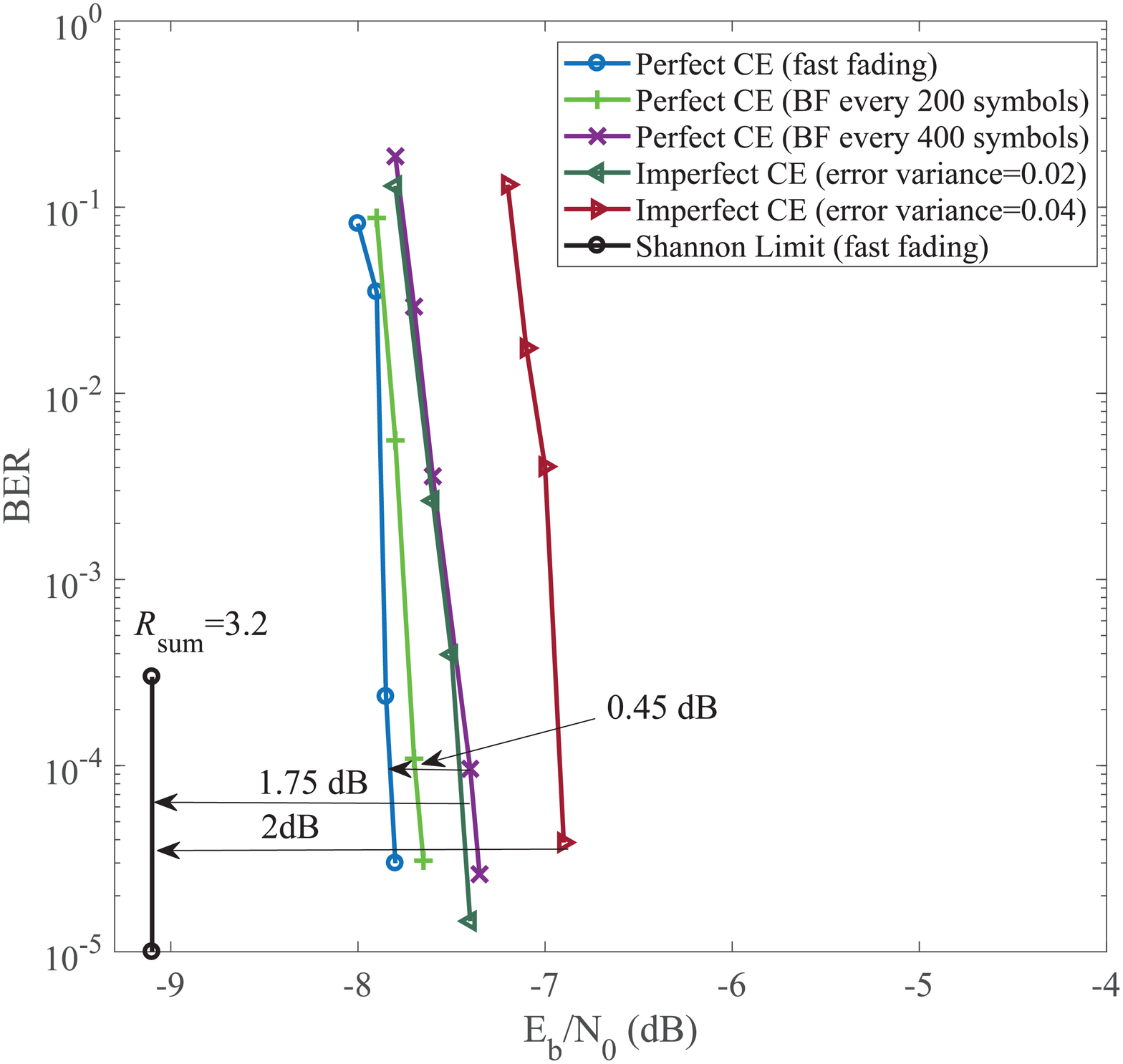}}
\caption{BER curves of the proposed systems over the MIMO-NOMA with full loading ($K=8, M=8$) and severe loading ($K=32, M=8$), in which the channels are fast fading, Block Fading (BF), and the imperfect CSI cases. In BF channels, channel fading parameters are unchanged for every $200$ and $400$ transmitted symbols. In the imperfect CSI cases, variances of channel estimated deviations are $0.02$ and $0.04$. CE denotes Channel Estimation.}
\label{ChanCode}
\end{figure*}

We consider the block fading channels, where the channel fading parameters remain unchanged for every 200 and 400 transmitted symbols of all users. Fig.~\ref{ChanCode} shows that BER curves at $10^{-4}$ of the proposed systems over block fading channels are about~$0.45$~dB from those of the fast fading cases, and are still within $1.75$~dB from the corresponding Shannon limits of the fast fading channels.

\subsubsection{Imperfect CSI}

In practical applications, channel estimation is difficult to be always estimated exactly. Therefore, we consider the fast fading channel and variances of estimated channel errors are $0.02$ and $0.04$. As shown in Fig.~\ref{ChanCode}, although imperfect channel estimations cause some performance losses, the gaps between BER curves at $10^{-4}$ of the proposed systems with imperfect CSI and the corresponding Shannon limits are within $2$~dB. This demonstrates that the proposed system is robust to the simulated channel conditions and imperfect channel estimations, which is favourable to the practical applications.

\section{Conclusion and Future Works}
In this paper, we proposed a practical MIMO-NOMA system for 5G communications, where transmitters and receiver were designed to achieve low complexity. The asymptotic EXIT analysis for the receiver consisting of LMMSE detector and message-passing decoders was provided to trace the statistical characteristics of estimated signals. Based on the asymptotic EXIT analysis, an MU-IRA coded MIMO-NOMA system was provided, whose implementation complexity was low and the asymptotic BER performances were within $0.2$~dB from the system capacity. Moreover, various numerical results were presented to validate the practicability and robustness of the proposed system. This implied that the proposed system would be an attractive solution for the MIMO-NOMA uplink in 5G communications.

There are two possible extensions for our work. One is finite-length code design, where the multi-user code distance analysis~\cite{SongMUCode} or scattered EXIT analysis~\cite{SEXIT} could be utilized for this task. Another extension is to further improving the iterative decoding threshold based on the spatial coupling techniques~\cite{SCTran,LiangISTC,Lingcl}.

\end{document}